\documentclass[preprint,journal]{vgtc}            

\onlineid{1059}

\vgtccategory{Research}

\vgtcpapertype{application/design study}

\newcommand{\sys}{\textsc{VisCanvas}\xspace}
\newcommand{\baseline}{\textsc{VisChat}\xspace}

\title{VisCanvas: A Node-Based Interface for Exploratory Visualization Authoring with LLMs}

\author{%
  \authororcid{Yuki Ueno}{0009-0008-9810-177X} 
  \and \authororcid{Bretho Danzy III}{0009-0003-2282-3833}
  \and \authororcid{Zhuojun Jiang}{0009-0008-0275-0661}
  \and \authororcid{Chris Bryan}{0000-0003-2430-815X} 
}

\authorfooter{
  \item
    {Yuki Ueno, Bretho Danzy III, Zhuojun Jiang, and Chris Bryan} are with Arizona State University. E-mail: \{yueno, bdanzy, zjian115, cbryan16\}@asu.edu
}

\abstract{%
  Visual data analysis involves both open-ended exploration and targeted question answering.
  Visualization authoring tools support this process by enabling users to create visualizations for these tasks.
  With the rise of large language models (LLMs), substantial effort has been devoted to developing visualization authoring tools that use natural language instructions.
  However, existing systems are typically based on a linear chat interface, which is not well suited to exploratory visual analysis workflows.
  In this paper, we introduce VisCanvas, a node-based interface for exploratory visualization authoring with LLMs.
  VisCanvas allows users to create, revise, branch, and merge visualizations in a non-linear way, enabling more efficient exploration of multiple analytical directions.
  We conducted a user study with 20 participants to evaluate the effectiveness of VisCanvas compared to a baseline chat-based interface.
  The results show that VisCanvas facilitates more diverse data interaction while maintaining performance levels (i.e., cognitive load and usability) that are indistinguishable from current prevailing methods.
  We then distill design principles for future AI-assisted visualization authoring environments.
  All supplemental materials required to reproduce the study are available at \url{https://osf.io/gsxhn}.
}

\teaser{
  \centering
    \includegraphics[width=\linewidth]{figs/teaser.pdf}
  \caption{%
    \sys supports exploratory visualization authoring through a node-based interface comprising: (a) a Context Menu for adding Text Input Nodes; (b) a Text Input Node for generating and updating visualizations, with (b1) a column-name autocompletion feature and (b2) a model selector; (c) a Visualization Node for viewing, interpreting, editing, and bookmarking results; (d) a Bookmark Panel for displaying bookmarked visualizations; (e) a Menu Node for creating new analysis paths via Modify, Duplicate, Merge, and Suggest operations; (f) an edge dropdown menu for extending connections via Fill Blank and Branch operations; and (g) a Data Explorer for inspecting datasets.
  }
  \label{fig:teaser}
}

\graphicspath{{figs/}{figures/}{pictures/}{images/}{./}} 

\usepackage{tabu}                      
\usepackage{booktabs}                  
\usepackage{lipsum}                    
\usepackage{mwe}                       
\usepackage{soul}

\usepackage{mathptmx}                  
\usepackage{float}
\usepackage{listings}
\usepackage{xcolor}
\colorlet{codeboxbg}{gray!10}
\colorlet{codeboxrule}{black!30}
\newlength{\codeboxframewd}
\usepackage[most]{tcolorbox}
\newtcolorbox{codeboxframe}{
    enhanced,
    breakable,
    colback=codeboxbg,
    colframe=codeboxrule,
    boxrule=\codeboxframewd,
    left=5pt,
    right=5pt,
    top=5pt,
    bottom=5pt,
    arc=0pt,
    boxsep=0pt
}
\usepackage{fancyvrb}
\usepackage{multirow}
\usepackage{caption}
\usepackage{graphicx}
\usepackage{tikz}
\usetikzlibrary{shapes}
\usepackage{pdfpages}

\usepackage[normalem]{ulem}

\definecolor{VisCanvasColor}{HTML}{6C8EBF} 
\definecolor{VisChatColor}{HTML}{D6B656}
\definecolor{magenta}{HTML}{D946EF}
\definecolor{text}{HTML}{EA580C}
\definecolor{visualization}{HTML}{0D9488}
\definecolor{suggestion}{HTML}{DB2777}
\definecolor{merge}{HTML}{7C3AED}
\definecolor{menunode}{HTML}{4F46E5}
\definecolor{duplicate}{HTML}{0D9488}
\definecolor{fillblank}{HTML}{475569}

\DeclareRobustCommand{\colorcircle}[1]{%
  \tikz[baseline=-0.5ex]{
    \fill[fill=#1] (0,0) circle (0.5em);
  }%
}

\DeclareRobustCommand{\iconcircle}[3]{%
  \tikz[baseline=-0.5ex]{
    \node[circle, fill={#1}, inner sep=0pt, minimum size=1em] {%
      \includegraphics[width=#2, height=#2, keepaspectratio]{#3}%
    };
  }%
}

\definecolor{beige}{rgb}{0.96, 0.96, 0.86}

\DefineVerbatimEnvironment{CodeBox}{Verbatim}{
    frame=single,
    framesep=5pt,
    fontsize=\small,
    rulecolor=\color{black!30},
    breaklines=true,
    breakanywhere=true
}

\lstdefinestyle{codebox}{
    frame=none,
    basicstyle=\ttfamily\small,
    gobble=4,
    moredelim=[s][\color{blue}]{state[}{]},
    emph={vega_lite_template, vega_lite_examples},
    emphstyle=\color{blue},
    breaklines=true,
    breakatwhitespace=true,
    breakindent=0pt,
    showstringspaces=false,
    columns=flexible,
    xleftmargin=0pt,
    resetmargins=true,
    aboveskip=0pt,
    belowskip=0pt
}

\lstdefinelanguage{json}{
    morestring=[b]",
    morekeywords={true,false,null},
    sensitive=true
}

\lstdefinestyle{jsoncode}{
    style=codebox,
    language=json,
    stringstyle=\color{red!60!black},
    keywordstyle=\color{blue},
    emph={},
    keepspaces=true
}

\newcommand{\OverviewMetaFont}{\small}

\newcommand{\OverviewParticipantLabel}{Participant ID}
\newcommand{\OverviewDatasetLabel}{Dataset}
\newcommand{\OverviewInterfaceLabel}{Interface}

\newcommand{\OverviewGraphic}[4]{%
  \begin{figure*}[t]
    \centering
    \setlength{\fboxsep}{6pt}
    \fbox{%
      \begin{minipage}{0.8\textwidth}
        \captionsetup{type=figure,labelformat=empty,labelsep=none}
        \centering
        \includegraphics[width=\columnwidth]{figs/overviews/pdf/#1}\par
        \vspace{0.5em}
        \caption[]{\OverviewMetaFont
          \textbf{\OverviewParticipantLabel:}~#2\quad
          \textbf{\OverviewDatasetLabel:}~#3\quad
          \textbf{\OverviewInterfaceLabel:}~#4}
      \end{minipage}
    }
  \end{figure*}
}

\begin{document}



\maketitle

\section{Introduction}

Today's proliferation of data has amplified the need for accessible visualization tools that lower the barrier to creating effective data representations~\cite{satyanarayan_critical_2020}. Such tools span levels of abstraction, ranging from formal grammars~\cite{bostock_d3_2011, satyanarayan_reactive_2016, satyanarayan_vegalite_2017, lyi_gosling_2022, hadleywickham_create_2016} to graphical interfaces~\cite{_business_, satyanarayan_lyra_2014, mendez_ivolver_2016, liu_data_2018, ren_charticulator_2019}. With recent advances in large language models (LLMs), researchers have begun exploring visualization authoring through natural-language instructions~\cite{dibia_lida_2023, han_chartllama_2023, maddigan_chat2vis_2023, tian_chartgpt_2025}. In these systems, users can iteratively define and refine visualizations by providing instructions through a chat or spoken-language interface in a conversational manner.

However, visual analysis is often inherently exploratory, as it involves both constructing visualizations and discovering knowledge through them.
For analysis-oriented tasks, exploration is driven by the need to uncover insights from data, ranging from answering targeted questions to identifying questions of interest through open-ended exploration~\cite{brehmer2013multi}.
In contrast, for design-oriented or storytelling contexts, the focus can center on the visual representation itself, such as comparing design alternatives and multiple representations to converge on an effective and expressive visualization~\cite{dow_parallel_2011}.


While authoring systems lower the barriers to creating visualizations, they often assume linear workflows~\cite{satyanarayan_critical_2020}  or depth-first exploration strategies~\cite{wongsuphasawat_voyager_2016}, and chat-based interaction paradigms further reinforce this linearity. 
This mismatch can become especially problematic during early-stage, analysis-oriented authoring, where users need to browse data, branch from partial findings, compare alternatives, and pursue multiple hypotheses. 
However, chat-based systems force users to externalize this exploratory reasoning through a single sequential interaction thread. 
As a result, it can become difficult to manage divergent analytical directions, compare alternatives side-by-side, and revisit the rationale behind prior decisions. 

This motivates the present work, which investigates how to support analysis-oriented exploratory visualization authoring within a human--LLM collaboration context. 
To this end, we introduce \sys, an AI-assisted visualization authoring system that adopts a node-based interaction paradigm. 
In \sys, each node represents a visualization instance together with its corresponding Vega-Lite~\cite{satyanarayan_vegalite_2017} specification, allowing users to explicitly branch, compare, revise, and merge alternative analytical directions. 
By externalizing the exploration space as a node graph, \sys enables users to maintain multiple threads or branches of analysis and exploration simultaneously and reason about analytical alternatives in parallel, rather than through a single linear conversational thread.

\sys is primarily positioned for early-stage analytical exploration and chart construction, where users iteratively generate, branch, compare, and refine Vega-Lite visualizations. Each visualization remains a reusable Vega-Lite specification, users can further refine encodings, labels, styling, and presentation details through specification editing or downstream authoring workflows.
It does this by supporting non-linear exploration during the phase in which users are still identifying promising questions, patterns, and analytical directions, and asking questions about the underlying dataset. 
To enable this, \sys combines a graph-centric frontend with a staged LLM pipeline that translates user instructions into analysis goals and Vega-Lite specifications while using self-reflection to detect and repair invalid outputs. 
The interface is designed around a graph-focused UI and UX to support chat-based authoring together with branching, merging, on-demand suggestions, and semantic zooming, allowing users to move fluidly between focused refinement and breadth-oriented exploration on an unlimited canvas.
In contrast to prior chat-based visualization tools that primarily reinforce linear workflows, \sys represents, to our knowledge, an early investigation into chat-based visualization authoring in an exploratory, non-linear manner.

We evaluate \sys through a controlled user study with 20 students. 
The results indicate that \sys promotes more exploratory authoring behaviors, supporting breadth-first exploration while maintaining a cognitive burden comparable to that of conversational interfaces (e.g., ChatGPT-like systems for visualization authoring). 
Participants also preferred \sys for open-ended exploratory scenarios because it better supported organizing thoughts, navigating the exploration space, parallel discovery, and comparing and merging visualizations to generate new insights.

In summary, this paper makes the following contributions:

\begin{itemize}
    \item We develop and study a node-based interaction paradigm that externalizes multi-strategy analytical reasoning beyond linear LLM conversations through operations such as branching, comparison, and merging.
    \item We present \sys, a node-based system for exploratory visualization authoring with LLMs. To support reproducibility, the tool's codebase is available at \url{https://github.com/svl-at-asu/VisCanvas}.
    \item We provide empirical evidence from a controlled user study (N=20) showing how node-based LLM interfaces shape exploratory authoring behaviors (study materials are available at \url{https://osf.io/gsxhn}). We also reflect on implications for future AI-assisted authoring environments.
\end{itemize}

\section{Related Work}

\subsection{Visualization Authoring Tools}

A variety of authoring approaches facilitate visualization creation. For instance, toolkits like D3.js~\cite{bostock_d3_2011} and Vega~\cite{satyanarayan_reactive_2016} offer full expressive power but require substantial programming expertise.
Other declarative systems, inspired by the \textit{Grammar of Graphics}~\cite{wilkinson_the_2012}, including Vega-Lite~\cite{satyanarayan_vegalite_2017}, Gosling~\cite{lyi_gosling_2022}, and ggplot2~\cite{hadleywickham_create_2016}, allow authors to easily specify a wide variety of visualizations using concise and uniform languages.
In addition to programming-based approaches, a range of interactive visualization authoring systems have been developed~\cite{satyanarayan_critical_2020}. 
For example, Tableau~\cite{_business_}, Power BI~\cite{_powerbi_}, and Polaris~\cite{stolte_polaris_2002} enable users to construct visualizations by mapping data fields to visual encoding channels, rather than relying on predefined templates~\cite{_microsoft_, _rawgraphs_}. 
Furthermore, computational notebooks such as Jupyter Notebook~\cite{kluyver_jupiter_2016} and Colab~\cite{_colab_} are common tools for seamlessly combining text, code, and visualizations in a document, which are typically presented in a linear manner~\cite{wang_supernova_2024}.

These tools allow authors to create visualizations efficiently, but they are generally well-suited to depth-first exploration strategies, in which users need to provide detailed specifications, hindering rapid exploration of multiple alternative analytical directions. 
As a result, broad exploration across such directions remains an open problem~\cite{wongsuphasawat_voyager_2016}. 
Specifically, Voyager~\cite{wongsuphasawat_voyager_2016} and Voyager~2~\cite{wongsuphasawat_voyager_2017} are particularly important examples of systems that support exploratory data analysis through visualization recommendation and faceted browsing.

While sharing similar goals of lowering authoring barriers, \sys differs in both scope and interaction model. 
In particular, compared with Voyager systems, \sys does not primarily support exploration through browsing system-recommended views over the data space. 
Instead, it focuses on human--LLM co-authoring, where users iteratively create, revise, branch, and combine visualizations. 
Our work focuses on early-stage, analysis-oriented exploratory authoring, where users are still forming questions and comparing possible analytical directions. 
To support this phase, \sys introduces a node-based interface that facilitates branching, comparison, and reuse of intermediate visualization states in human--LLM collaboration.

\subsection{Using LLMs for Visualization Construction}

Natural language interfaces have been extensively explored across various stages of visualization workflows~\cite{shen_natural_2023}. 
Recent advances in LLMs further enable the translation of textual descriptions into specifications or code~\cite{wang_visualization_2025, hutchinson_foundation_2025}. 
Systems such as LIDA~\cite{dibia_lida_2023}, ChartLlama~\cite{han_chartllama_2023}, Chat2VIS~\cite{maddigan_chat2vis_2023}, PlotGen~\cite{goswami_plotgen_2025}, and CoDA~\cite{chen_coda_2025} leverage LLMs to generate Python code for data visualization. 
Other systems, including DynaVis~\cite{vaithilingam_dynavis_2024}, ChartGPT~\cite{tian_chartgpt_2025}, and VisPilot~\cite{wen_exploring_2026}, generate Vega-Lite specifications; meanwhile InterChat~\cite{chen_interchat_2025} produces D3.js code through interaction-augmented instructions~\cite{shen_prompting_2025}. 
More closely related to exploratory visual analysis, AI Threads~\cite{hong_data_2025} is an LLM-based prototype that combines a Main Chat and Thread panel to support branching dialogue.
While these systems lower the barrier to entry, they largely rely on linear, chat-based human--LLM interaction, where users prompt the LLM and the visualization is iteratively updated.
Such interaction models can be seen as aligning with linear workflow assumptions~\cite{satyanarayan_critical_2020} or depth-first exploration strategies~\cite{wongsuphasawat_voyager_2016}, similar to those found in many visualization authoring tools.
This linear workflow is particularly limiting for analysis-oriented exploratory authoring, in which users often need to branch from intermediate results, compare alternative views, and revisit partially developed ideas. 
Without explicit structural mechanisms for relating these alternatives, users struggle to maintain a mental model of the evolving authoring process. 

To address the limitations of linear interaction histories, some systems have explored recording user interactions and LLM responses graphically~\cite{heer_graphical_2008}. VisPilot~\cite{wen_exploring_2026} integrates an Authoring Flow that records text/visual prompts and LLM responses as a chronological process view, with generated visualizations shown to help users review how an authoring task was completed; however, these recorded states primarily support retrospective review rather than reuse as actionable starting points for forking or recombining analysis.
Urbanite~\cite{moreira_urbanite_2026} goes further by using LLM-assisted executable dataflows for urban visual analytics, where heterogeneous nodes encode data loading, transformation, and visualization modules connected by data or interaction dependencies.
This provides strong support for constructing and inspecting analytics pipelines, but its exploration is centered on workflow modules and dataflow versions; users typically explore alternatives by modifying pipelines rather than directly manipulating lightweight candidate charts.
\sys addresses this gap by making generated visualization states the primary exploration units: each Visualization Node is an actionable Vega-Lite state that users can revisit, revise, duplicate, branch from, compare, and merge during early-stage analysis.

\subsection{Provenance for Visualization Creation}

Extensive visualization research has been devoted to supporting provenance~\cite{ragan_characterizing_2016, xu_survey_2020}. 
Ragan et al.~\cite{ragan_characterizing_2016} surveyed provenance systems and characterized prior work according to the types of provenance information captured and their purposes. 
Similarly, Xu et al.~\cite{xu_survey_2020} reviewed provenance research using a Why--What--How framework. 
Prior studies have shown that provenance information supports users in reviewing past actions, comparing alternatives, and reconstructing analytic processes, thereby facilitating sensemaking, discovery, and creative reasoning~\cite{xu_survey_2020, shneiderman_creativity_2007}. 
Accordingly, provenance systems have emphasized retrospective understanding and explanation of analysis workflows.

From the perspective of visualization creation, many systems record intermediate visual outputs produced during analysis~\cite{ragan_characterizing_2016}. 
A number of visual analytics tools (e.g.,~\cite{ma_image_1999, callahan_vistrails_2006, dunne_graphtrail_2012, javed_explates_2013, dobos_3d_2014, yu_visflow_2017, ulbrich_smolboxes_2023}) employ graphical histories~\cite{heer_graphical_2008} to represent the evolution of visualization states, enabling users to save, revisit, and review prior visualizations. 
In particular, systems such as Image Graphs~\cite{ma_image_1999}, GraphTrail~\cite{dunne_graphtrail_2012}, ExPlates~\cite{javed_explates_2013}, VisFlow~\cite{yu_visflow_2017}, and sMolBoxes~\cite{ulbrich_smolboxes_2023} adopt graph-based interfaces in which each node represents a visualization state. 
Building on this line of work, FlowSense~\cite{yu_flowsense_2020} integrates natural language processing into VisFlow, allowing users to construct and modify visualizations through conversational input. 
While natural language interfaces reduce learning overhead, they still require users to specify precise computational instructions due to limitations in language understanding and inference.

More recently, provenance-inspired designs have been explored in the HCI community to help users organize ideas and iteratively refine instructions when working with generative models~\cite{kim_metaphorian_2023, angert_spellburst_2023, suh_sensecape_2023, liu_how_2024, guo_prompthis_2025}. 
These systems demonstrate the value of representing generative outputs as nodes, enabling users to reflect on, compare, and manage multiple ideas. 
For instance, Spellburst~\cite{angert_spellburst_2023} closely aligns with our interaction model by representing LLM-generated artifacts as nodes that users can branch and merge. Unlike Spellburst, which focuses on creative-coding variants, \sys treats nodes as data-grounded Vega-Lite visualization states tied to analysis goals, enabling visualization-specific suggestions, comparison, and merging.

Our work extends this line of research to AI-supported visualization authoring, where node-based provenance structures must support not only idea management but also data-grounded analytical exploration.
By foregrounding prior visualization states as starting points for further exploration, our work advances provenance and authoring-focused research toward analysis-oriented exploratory visualization authoring.

\section{Design Considerations}
\label{sec:design_challenges_goals}

To develop a design that achieves our goal of analysis-oriented exploratory visualization authoring within a human-LLM collaborative context, we present a set of design considerations.
These were distilled from a review and synthesis of prior work in AI-assisted data analysis~\cite{gu_how_2024}, visualization authoring with LLMs~\cite{vaithilingam_dynavis_2024, tian_chartgpt_2025}, exploratory search and analysis~\cite{white_exploratory_2009, wongsuphasawat_voyager_2016}, and provenance-inspired design~\cite{xu_survey_2020, park_storyfacets_2022, liu_how_2024, angert_spellburst_2023}.

\textbf{C1: Exploration structure at multiple levels of granularity.}
In analysis-oriented visualization authoring, users often pursue multiple partial hypotheses, revisit prior results, and compare alternative directions. Explicitly showing the structure of these states (including their branching) can support overview and recall~\cite{ragan_characterizing_2016, xu_survey_2020}, however linear authoring tools are constrained to only a sequential view of such history. At the same time, prior work such as StoryFacets shows that branching history views can become overwhelming as analyses grow~\cite{park_storyfacets_2022}. One potential solution is \textit{providing explicit views of the exploration structure at multiple levels of granularity (i.e., both local and global detail levels) that emphasize the exploratory and branching structure of analysis while allowing users to focus on level-of-detail specific information}.

\textbf{C2: On demand recommendations to assist in data exploration.}
Exploratory visual analysis requires users not only to refine existing ideas but also to ideate and explore promising next steps. Prior work such as Voyager showed that visualization recommendations can effectively support breadth-oriented exploration by helping users survey possible directions and encounter views they might not have specified on their own~\cite{wongsuphasawat_voyager_2016}. At the same time, such suggestions can be detrimental if they impose too much cognitive overhead, are excessive, poorly timed, or remove the user's locus of control~\cite{zhou2023design}. Therefore, \textit{suggestions should be lightweight, context-aware, and on demand (i.e., triggered by users)}.

\textbf{C3: Merging and branching of ideas.}
As mentioned for \textbf{C1}, analysis-oriented authoring rarely proceeds as a sequential chain of revisions. Users may wish to go back and fork from an intermediate chart, preserve one promising path while testing another, or synthesize insights from multiple views into a new comparative visualization. Prior (non-visualization) provenance-inspired and generative-authoring systems such as Spellburst and CoQuest have shown the value of branching and combining intermediate outputs for creative exploration in tasks such as programming and brainstorming~\cite{angert_spellburst_2023, liu_how_2024}. \textit{Likewise, visualization-focused design tools should treat visualization states as reusable objects that users can branch from, duplicate, and merge, so that divergent analytical directions can be explored and later brought back into relation with one another.}

\section{System Design and Implementation}

\sys is a visualization authoring tool that supports non-linear exploration via human--LLM collaboration, with an emphasis on supporting rapid visualization authoring that is common in the early stages of data analysis.
Below, we describe the interface features and system implementation details of \sys (including how various features are designed to support \textbf{C1--C3}); evaluations are presented in \cref{sec:evaluation}. 
To support reproducibility, the system's codebase is available at \url{https://github.com/svl-at-asu/VisCanvas}.

\subsection{User Interface \& Features}
The \sys interface (see \cref{fig:teaser} for the entire UI) is presented as a set of components on a zoomable and pannable canvas (like Miro or Figma). Users can create and manipulate different types of nodes, which are connected by edges to indicate the analysis workflow. Specifically, \sys supports four types of nodes (Text Input Node, Visualization Node, Merge Node, Menu Node). \sys also supports edge-specific operations and additional usability features such as semantic level-of-detail zooming and a data explorer panel for reviewing the dataset. Below, we describe these components and features and note how they support the design considerations introduced in \cref{sec:design_challenges_goals}.

\textbf{Text Input Nodes:}
\label{sec:text_input_node}
Through the Context Menu (\cref{fig:teaser}-a), opened by right-clicking, users can add a Text Input Node to start a new graph. In this node, they can input natural language instructions to generate Vega-Lite specifications (\cref{fig:teaser}-b) or modify an existing Vega-Lite specification (\cref{fig:menu_node_output}-a).
The text input field supports autocompletion of the column names in the dataset (\cref{fig:teaser}-b1), and users can also select the desired LLM for generating the Vega-Lite specifications (\cref{fig:teaser}-b2). For testing, \texttt{gpt-4.1-nano}\footnote{\small https://platform.openai.com/docs/models/gpt-4.1-nano} and \texttt{gpt-5-nano}\footnote{\small https://platform.openai.com/docs/models/gpt-5-nano} were used,
though the system is easily extensible to accommodate additional models.
When a user submits a prompt from a Text Input Node, the system creates a new Visualization Node and connects it to the Text Input Node via an edge if one does not already exist. Upon receiving the backend response, the system updates the \textit{Visualization Node's} specification; this update then cascades through  all connected downstream nodes, enabling the automatic revision of an entire visualization chain.

\begin{figure}[h]
    \centering
        \includegraphics[width=\columnwidth]{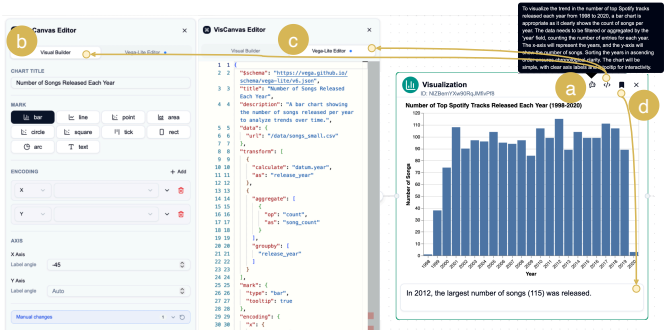}
    \caption{Visualization Node: (a) Robot icon shows the rationale for the LLM-generated results; (b) Visual Builder and (c) Vega-Lite Editor are provided for editing the Vega-Lite specification; (d) Bookmark icon allows users to bookmark the visualization for later reference.}
    \label{fig:vis_node}
\end{figure}

\begin{figure}[ht]
    \centering
        \includegraphics[width=\columnwidth]{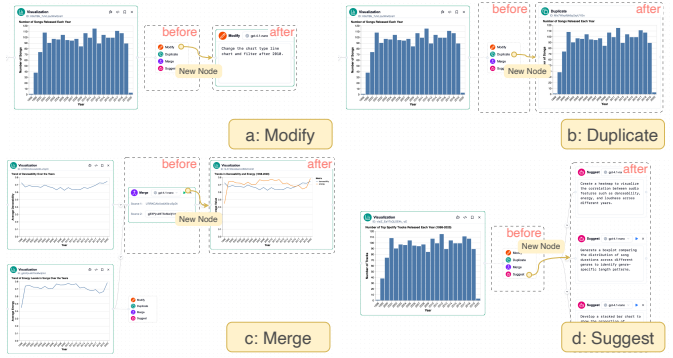}
    \caption{(a) \textit{Modify} enables users to generate visualizations via natural language; (b) \textit{Duplicate} clones the parent visualization; (c) \textit{Merge} combines two distinct data views; (d) \textit{Suggest} recommends 3--5 LLM-generated prompts to users.}
    \label{fig:menu_node_output}
\end{figure}

\textbf{Visualization Nodes:}
Vega-Lite charts are rendered in Visualization Nodes (\cref{fig:teaser}-c). Each node renders a standard Vega-Lite specification using Vega-Lite's renderer, enabling it to display interactive or composite visualizations, including basic interactions such as filtering and panning and zooming, as well as multiple linked views within a single node.
In addition, Visualization Nodes offer several functionalities to help users understand, refine, and interact with the generated visualizations. For example, hovering over an icon reveals the rationale for an LLM-generated visualization (\cref{fig:vis_node}-a), clarifying the system's design choices~\cite{gu_how_2024}.
Users can further refine the visualization through two editing interfaces: a \textit{Visual Builder} (\cref{fig:vis_node}-b), a widget-based GUI for modifying the specification without writing code, and a \textit{Vega-Lite Editor} (\cref{fig:vis_node}-c) for directly editing the specification as raw text.
These GUI editors are intended to complement natural language instructions by offering fine-grained control, immediate visual feedback, and quick undo/redo of edits~\cite{vaithilingam_dynavis_2024}.
Users can also bookmark a visualization and leave a note for later reference (\cref{fig:vis_node}-d); bookmarked visualizations are displayed in a Bookmark Panel (\cref{fig:teaser}-d).

\textbf{Menu Nodes [C2, C3]:}
Menu Nodes provide four operations for refining and extending the exploration (\cref{fig:teaser}-e).
\textit{Modify} adds a new Text Input Node as a child of a parent Visualization Node, allowing users to issue new instructions that further explore or refine the existing visualization (\cref{fig:menu_node_output}-a).
\textit{Duplicate} creates a copy of the parent Visualization Node as a new independent node in the graph, allowing users to edit the visualization while keeping the original visualization (\cref{fig:menu_node_output}-b).
\textit{Merge} adds a Merge Node and allows users to combine two Visualization Nodes (labeled Source~1 and~2) into a single comparative visualization (\cref{fig:menu_node_output}-c): Source~1 is automatically set to the parent node, while Source~2 can be specified by entering a node ID or by activating cursor mode and clicking the target node (\textbf{C3}).
The \textit{Merge} operation is prompt-guided rather than based on a hand-coded lookup table over source chart types. The prompt asks the LLM to identify shared attributes, align shared axes when present, and generate a layered or dual-axis chart to support direct comparison, with layered charts limited to at most two layers for readability; otherwise, it constructs a related view, such as a repeated chart, that reveals relationships or contrasts between the two sources (see Appx.~A.2).
\textit{Suggest} prompts the system to generate 3--5 candidate visualization instructions as next analysis steps (\cref{fig:menu_node_output}-d), grounded in the analysis goals (a by-product of Vega-Lite specification generation discussed in \cref{sec:backend}) accumulated along the path from the root to the current node; these suggestions are instantiated as persistent Text Input Nodes on the canvas, so users can return to them and adopt them later (\textbf{C2}).

\begin{figure}[t]
    \centering
        \includegraphics[width=\columnwidth]{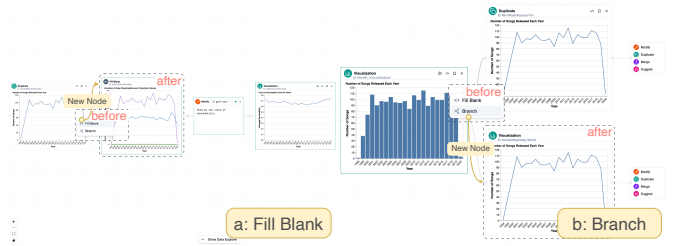}
    \caption{(a) \textit{Fill Blank} inserts an intermediate visualization between two connected nodes; (b) \textit{Branch} creates a Menu Node or Visualization Node based on the edge's target node.}
    \label{fig:fill_blank_branch}
\end{figure}

\textbf{Edge Dropdown Menus [C3]:}
Edge Dropdown Menus provide two operations for extending the exploration (\cref{fig:teaser}-f).
\textit{Fill Blank} generates an intermediate Visualization Node that bridges the analytical gap between two connected Visualization Nodes (\cref{fig:fill_blank_branch}-a).
Like the \textit{Merge} operation, this operation is prompt-guided: the prompt asks the LLM to identify shared or related attributes, differences in analytical focus, and potential causal or correlational relationships between the two endpoint visualizations, then generate a Vega-Lite specification that makes the reasoning between them more explicit (see Appx.~A.2).
The generated node is inserted between the two existing nodes as a new Visualization Node.
\textit{Branch} handles on the target node of the selected edge: if the target is a Visualization Node, it creates a new independent copy; otherwise, it inserts a new Menu Node (\cref{fig:fill_blank_branch}-b). In either case, users can explore an alternative analysis direction without modifying the original graph structure.

\begin{figure}[ht]
    \centering
        \includegraphics[width=\columnwidth]{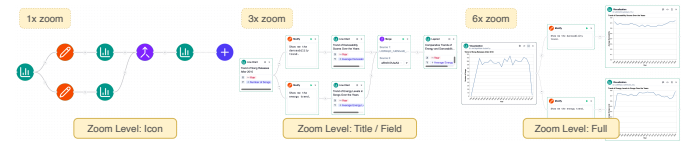}
    \caption{Semantic Zoom: Users can navigate large workflows by adjusting detail by zoom level.}
    \label{fig:semantic_zoom}
\end{figure}

\textbf{Semantic Zoom [C1]:}
To support navigation as the exploration graph grows in size, \sys employs a semantic zoom feature, which adapts the level of detail shown for each node based on the current zoom level (\cref{fig:semantic_zoom}).
When zoomed out, all nodes are rendered as compact icons, providing a high-level overview of the graph structure (\textit{Icon} zoom level).
Zooming in changes nodes to the \textit{Title/Field} zoom level, where Visualization Nodes display their title alongside the field names assigned to each encoding, allowing users to quickly identify the content of each chart without rendering the full visualization.
At the \textit{Full} zoom level, Visualization Nodes render the complete Vega-Lite chart.
Text Input Nodes, Merge Nodes, and Menu Nodes maintain the same appearance at both the \textit{Title/Field} and \textit{Full} levels.

\begin{figure}[ht]
    \centering
        \includegraphics[width=\linewidth]{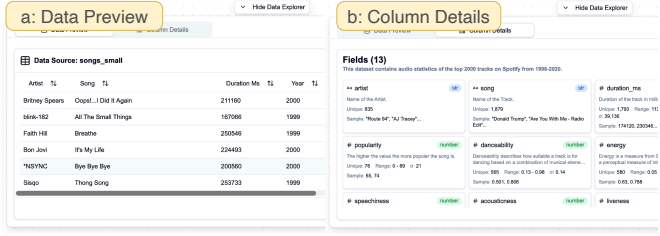}
    \caption{Data Explorer: Users can explore the datasets by (a) data preview and (b) column details.}
    \label{fig:data_explorer}
\end{figure}

\textbf{Data Explorer:}
The Data Explorer panel is accessible via the \textit{Show Data Explorer} button at the bottom of the canvas screen (\cref{fig:teaser}-g), and comprises a \textit{Data Preview} and \textit{Column Details} pair of tabs.
\textit{Data Preview} (\cref{fig:data_explorer}-a) displays the dataset in a tabular format, while the \textit{Column Details} (\cref{fig:data_explorer}-b) shows metadata for each column, including its description, data type, and data distribution.

\subsection{Implementation Details}
\subsubsection{Frontend}

The \sys frontend is implemented using a combination of React, Reactflow\footnote{\small https://reactflow.dev/}, Zustand\footnote{\small https://zustand.docs.pmnd.rs/}, and elkjs\footnote{\small https://github.com/kieler/elkjs}. 
React is the foundation for our user interfaces, while Zustand is used for state management of nodes and edges of the graph.
For the graph visualization, we use Reactflow to build the node-based interface for exploratory visualization authoring.
We also use elkjs to automatically arrange the graph in a hierarchical manner to improve the readability of the graph.
Within Reactflow, rendering Vega-Lite visualizations as SVG caused repeated re-rendering during node movement and led to sluggish interactions, so we instead used canvas-based rendering to improve responsiveness. 
To make the editing experience more intuitive, the system updates visualizations in real-time as users modify specifications via the Visual Builder or Vega-Lite Editor. The automatic layout is triggered whenever new nodes are added or the graph dimensions change. An example of the JSON objects used to manage node and edge states is provided in Appx.~C.

\subsubsection{Backend}
\label{sec:backend}

The \sys backend is implemented using FastAPI\footnote{\small https://fastapi.tiangolo.com/} to orchestrate core system functionalities, including visualization specification generation and analysis goal suggestion.
For the visualization specification generation and analysis goal suggestion, we utilize LangGraph\footnote{\small https://www.langchain.com/langgraph} to coordinate multiple LLM-based agents.
An overview of the system flow is shown in Fig.~11-a,b
in Appx.~D; we detail the steps below.

\textbf{Visualization Specification Generation:}
When a user submits either a natural language instruction via a Text Input Node or a predefined visualization instruction for the \textit{Merge} and \textit{Fill Blank} operations, the pipeline takes the instruction and any existing Vega-Lite specifications from upstream nodes as input.
Because each operation is handled as a separate pipeline invocation, \sys does not maintain a persistent LLM session or implicit chat history across interactions. Instead, each operation issues a new LLM call with the relevant graph context explicitly supplied in the prompt.
Accordingly, the current pipeline is designed for user instructions that can be interpreted as requests to produce or revise a Vega-Lite specification, rather than for open-ended multi-turn dialogue in which the system asks follow-up questions before generating a visualization.
This pipeline comprises three LLM-powered modules orchestrated using LangGraph (Fig.~11-c):
\textit{Data Summarizer}, \textit{Analysis Goal Generator}, and \textit{Vega-Lite Spec Generator}.

Based on empirical review during development and testing, we employ \texttt{gpt-4.1-nano} for the \textit{Data Summarizer} and \textit{Analysis Goal Generator} modules, and either \texttt{gpt-4.1-nano} or \texttt{gpt-5-nano} for the \textit{Vega-Lite Spec Generator} module; users can select the model for each Text Input Node as described in \cref{sec:text_input_node} (see also \cref{sec:performance_results} for an analysis of these models' performance during our evaluations).

The \textit{Data Summarizer} module analyzes the uploaded dataset and generates a compact, information-dense summary to serve as grounding context for subsequent LLM modules.
Although passing an entire dataset directly to the LLM is feasible with recent long-context models, doing so would significantly increase response latency and incur unnecessary cost, particularly as datasets grow in size and dimensionality.
To this end, we adopt the Data Summarizer from LIDA~\cite{dibia_lida_2023}, which extracts dataset properties including atomic types, general statistics, and representative values, enriched with LLM-generated semantic descriptions for the dataset and fields.
Note that the Data Summarizer is invoked only once when a dataset is first uploaded; thereafter, the summary is cached and reused across all subsequent steps.

The \textit{Analysis Goal Generator} module produces an intermediate analysis goal that bridges the user's natural language instruction and the final visualization specification.
Given the data summary, the user's visualization instruction, and any existing Vega-Lite specifications from upstream nodes (if present), the module generates a concise analysis goal that captures the analytical intent of the request (see Appx.~A.1
for the full prompt).

Finally, the \textit{Vega-Lite Spec Generator} module transforms the analysis goal into a valid Vega-Lite specification.
Rather than selecting or scoring chart-type-specific templates, the module uses a single generic Vega-Lite template with \texttt{<stub>} fields, which provides a flexible structure for generating different chart designs.
Since LLMs have been shown to lack sufficient knowledge of Vega-Lite grammar~\cite{wang_visualization_2025}, this module also provides curated Vega-Lite specification examples in the prompt as few-shot guidance toward well-formed output structures.
It also outputs a short \textit{rationale} explaining the visualization design to help users understand the intent behind the LLM's output.
To improve reliability, the module incorporates a self-reflection mechanism: if the generated specification fails to compile, the module captures the error message and re-prompts the LLM to diagnose and correct the issue, iterating until a valid specification is produced (see Appx.~A.2
for the full prompt).

The frontend displays the generated Vega-Lite specification and rationale to the user in a Visualization Node. In addition, the frontend stores the analysis goal in each Visualization Node for the analysis goal suggestion.

\textbf{Analysis Goal Suggestion:}
When a user selects the \textit{Suggest} option on a Menu Node, the system collects the analysis goals stored in the Visualization Nodes along the path from the root to the current node.
These accumulated goals, together with the cached data summary, are provided to \texttt{gpt-4.1-nano}, which generates 3--5 candidate visualization instructions as relevant next analysis steps based on the context of the preceding exploration path (see Appx.~B for the full prompt).
The suggested instructions are then presented as Text Input Nodes, which users can directly adopt to continue their exploration.






\section{Evaluation}
\label{sec:evaluation}

\subsection{User Study Design}
\label{sec:user_study_design}

To empirically evaluate \sys, we conducted a within-subjects study with 20 participants. During each session, participants performed two sets of exploratory analysis tasks requiring them to generate Vega-Lite visualizations and extract distinct insights from the data. For each participant, one task was conducted using \sys, and the other using a baseline, chat-based interface.

\begin{figure}[h]
    \centering
    \includegraphics[width=\linewidth]{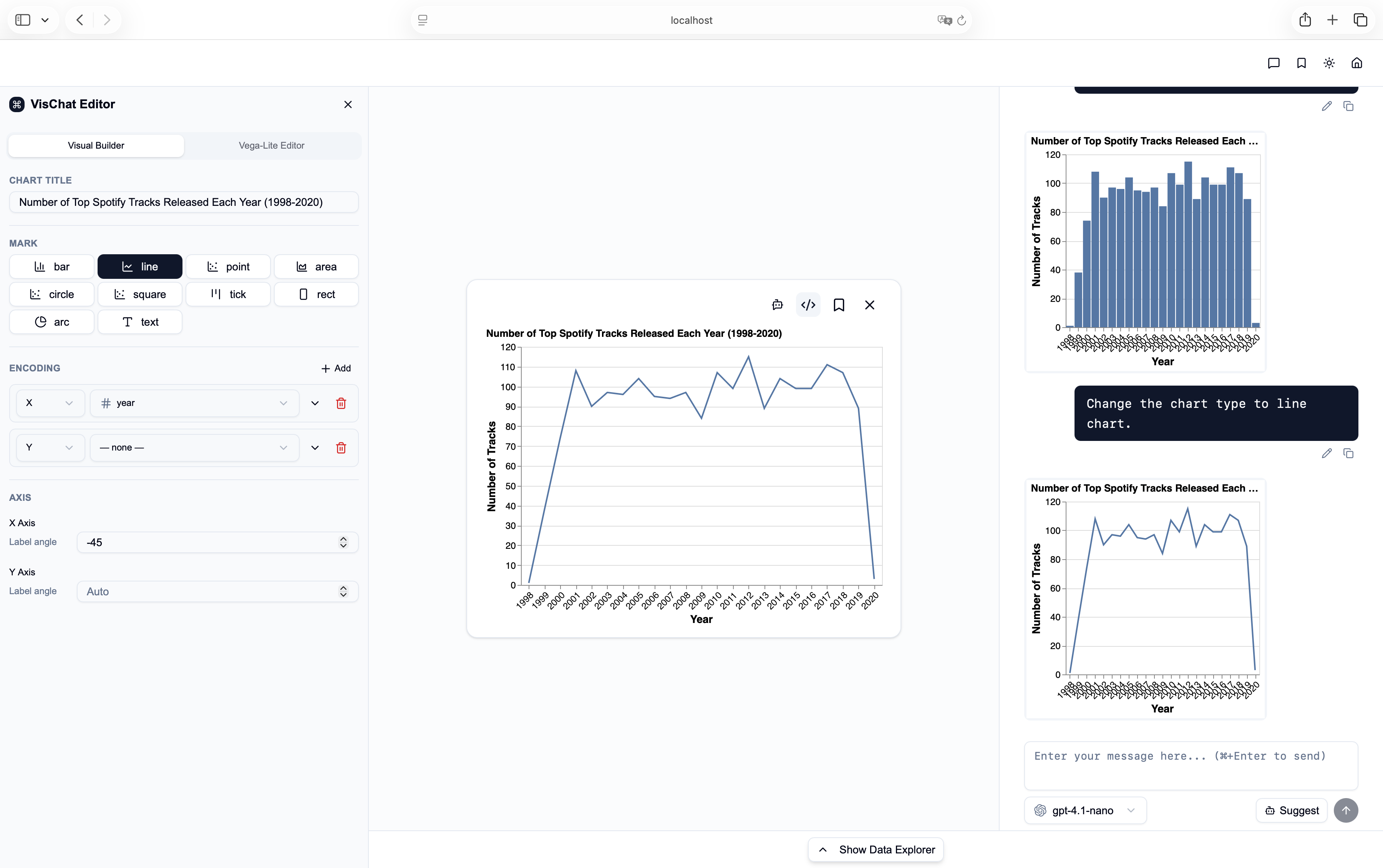}
    \caption{\baseline, a controlled chat-based authoring baseline that preserves the familiar linear layout of commercial LLM chat interfaces while including suggestion and visualization-editing capabilities. The right panel shows the chart and chat history; the middle and left panels show the Visual Builder for editing the current chart design and its properties.}
    \label{fig:vischat}
\end{figure}

\textbf{Experimental Baseline.}
We developed \baseline as a controlled chat-based baseline for testing against \sys. 
Rather than a literal replica of commercial systems such as ChatGPT or Gemini, \baseline preserves their linear chat layout while matching key capabilities available in \sys. 
Specifically, \baseline includes the same \textit{Suggest} backend, Visual Builder, and Vega-Lite Editor as \sys, controlling for access to LLM-generated next-step recommendations and low-level chart refinement. 
The two interfaces expose these capabilities through interface-native mechanisms: suggestions appear as editable prompt candidates in \baseline, whereas they become Text Input Nodes in \sys. 
Thus, the primary comparison focuses on the linear chat interface versus the node-based spatial interface.

\textbf{Research Questions.} To help scaffold our study's analyses and interpretations, we posed the following research questions:

\begin{itemize}[noitemsep]
\item \textbf{RQ1:} Does a node-based interface (i.e., \sys) make exploratory visual analysis easier or harder than a linear chat-based interface (i.e., \baseline)?
\item \textbf{RQ2:} Do users prefer a node-based interface over a linear chat-based interface for exploratory visual analysis?
\item \textbf{RQ3:} What features or interactions in \sys are considered beneficial or detrimental to exploratory visual analysis?
\item \textbf{RQ4:} What strategies do users employ during exploratory visual analysis with a node-based vs. a linear chat-based interface?
\item \textbf{RQ5:} Is \sys's backend architecture and system flow sufficiently responsive and performant for interactive use?
\end{itemize}

\textbf{Datasets.}
To evaluate \sys across varied contexts, we chose three datasets based on their size, complexity, and high potential for participant engagement: an \emph{Automobile} dataset~\footnote{\small https://www.kaggle.com/datasets/athirags/car-data} (used exclusively for warm-up exercises), a \emph{Spotify} dataset~\footnote{\small https://www.kaggle.com/datasets/paradisejoy/top-hits-spotify-from-20002019}, and a \emph{Social Media \& Mental Health} dataset~\footnote{\small https://www.kaggle.com/code/nuhmanpk/student-social-media-and-relationship-analysis}. 
The latter two were used for the main tasks, with their order and the assigned interface (\sys vs. \baseline) counterbalanced to mitigate order or data preference effects.
The \emph{Spotify} dataset has 2,000 records and 13 variables (3 nominal, 1 temporal, 9 quantitative).
The \emph{Social Media \& Mental Health} dataset has 705 records and 13 variables (6 nominal, 7 quantitative).
One participant reported that they had seen the \emph{Spotify} dataset before the study (their activities on this dataset were not unusual compared to other participants, so we included their results).
All datasets are available in our supplemental \href{https://osf.io/gsxhn}{OSF repository}.

\textbf{Study Procedure.}
The study was conducted in real time via Zoom to allow participants to work in their preferred environment. Sessions were expected to last 90 minutes; they averaged 86.8 minutes and ranged from 66 to 107 minutes. After an introductory phase, participants completed two conditions (in random order). For each condition, they watched a tutorial video, completed a warm-up exercise, performed the main task, and filled out a post-task survey. An optional break was provided between the two conditions, and a final post-study survey was administered at the end.
This study was classified as non-human-subjects research by our university's Institutional Review Board (IRB); nevertheless, we obtained informed consent from all participants.

\textit{Introduction and Consent}: Participants provided informed consent prior to the session. At the start of the session, the facilitator introduced the study's purpose and provided an overview of the interface features (1--2 minutes), followed by a pre-study demographic survey (5 minutes).

\textit{Tutorial Video and Warm-up (per condition)}: To minimize facilitator effects, participants viewed a prerecorded demonstration video (approx. 2 minutes) for the assigned condition, followed by a hands-on warm-up period (5--7 minutes) to explore the features and interaction patterns of the interface.

\textit{Main Task (per condition)}: Each participant conducted a free-form exploratory analysis of the assigned dataset (limited to 20 minutes), aiming to produce visualizations and record at least four bookmarked insights. To promote engagement with the analysis, participants were asked to give a brief oral report of their findings at the end of each condition, referencing the visualizations they had bookmarked.
We did not divide the exploratory analysis task into subtasks (i.e., exploration \& question answering), preserving the natural exploratory process~\cite{battle_characterizing_2019}.

\textit{Post-Task Survey (per condition)}: After completing the main task, participants filled out a short questionnaire that included the NASA-TLX to assess workload with additional questions to evaluate usability.

\textit{Break}: An optional break (3--5 minutes) was provided between the two conditions to mitigate fatigue. Participants then proceeded to the second condition using the alternate interface and dataset (i.e., if they first used \sys with the Spotify dataset, they next used VisChat with the Social Media dataset).

\textit{Post-Study Survey and Debrief}: After completing both conditions, participants filled out a final post-study survey (5--10 minutes) covering overall interface preferences and open-ended feedback. Immediately after the survey, we asked them whether they had any additional feedback or questions (1--2 minutes).

\textbf{Participants.}
We recruited 20 graduate computer science students (13 males, 7 females; age: $\mu = 23.25, \sigma =1.52$) from Arizona State University (P1--P20). While all participants were familiar with data visualization tools like D3.js, their experience with other technologies varied (see Appx.~E
for full details). 
Most were frequent users of LLMs (90\% at least weekly) and possessed high self-reported proficiency in data visualization (85\%) and data analysis (75\%). However, this expertise did not fully extend to the intersection of these fields; only 40\% used LLMs specifically for visualization tasks on a weekly basis. Furthermore, despite \sys utilizing a canvas-based interaction paradigm similar to Miro or Figma, participants reported relatively little experience with such whiteboard-style formats (16\% at least weekly).

\subsection{Analysis \& Results}
\label{sec:user_study_results}

During the study, we collected both quantitative and qualitative information, including task- and usage-related information, survey responses, open-ended freeform feedback, and system logging of participant actions. We report and discuss these results in the following sections (with the $p$-values from the Wilcoxon signed-rank tests provided in parentheses where applicable).

\subsubsection{Assessing Task Success and Perceived Difficulty (RQ1)}

As an initial sanity check, we first verified that participants could complete the \textit{Main Task} stages using the interfaces.
For both interfaces, all participants successfully created at least four visualizations, which was considered ``success'' in our study design.

In terms of how ``easy'' participants considered it was to complete each task stage, in the \textit{Post-Task Surveys} participants reported comparable levels of task completion ease for both \sys ($\mu = 6, \sigma = 1.34$) and \baseline ($\mu = 6.05, \sigma = 1.12$), with no significant difference between the two conditions ($p$-value $=0.89$). 
Similarly, participants expressed comparable confidence in the thoroughness of their analysis in \sys ($\mu = 5.8, \sigma = 0.93$) and \baseline ($\mu = 5.75, \sigma = 1.04$), again with no significant difference ({$p$-value $= 0.80$}) (see \cref{fig:nasa_tlx_combined_boxplot}).
These results are not surprising, given that participants performed an open-ended task that was capped at 20 minutes, with a relatively small number of required visualization bookmarks.
To better understand how each interface promoted different exploration strategies, we subsequently conducted a more detailed analysis of user authoring behavior (see \cref{sec:user_strategies}).

\subsubsection{Workload and Feature Assessments (RQ1)}
\begin{figure}[h]
    \centering
    \includegraphics[width=\linewidth]{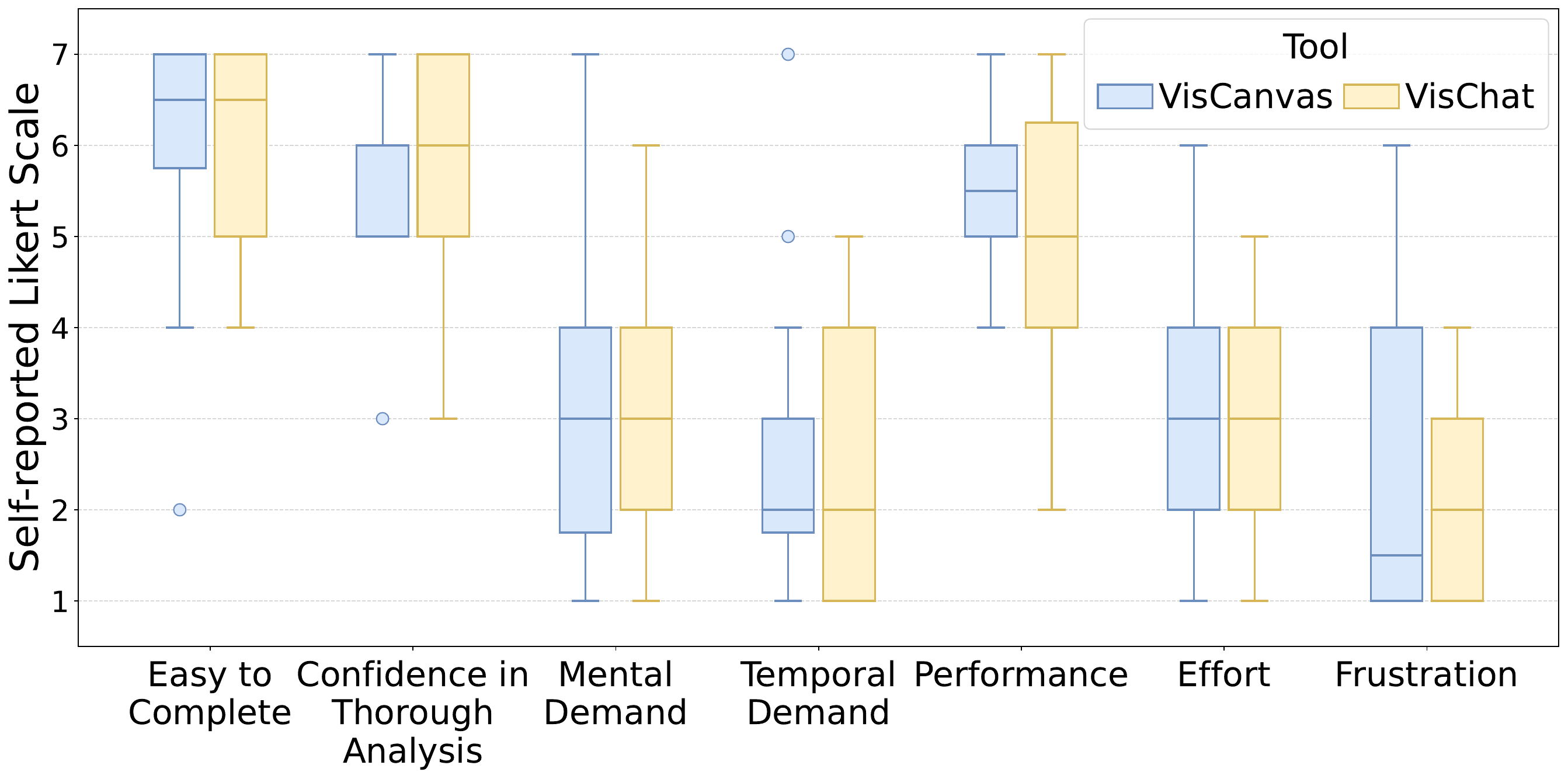}
    \caption{Participants' self-reported scores for NASA TLX questions, ease of completing the task, and confidence in their analysis thoroughness.}
    \label{fig:nasa_tlx_combined_boxplot}
\end{figure}

In each \textit{Post-Task Survey}, participants also completed NASA-TLX ratings  and a set of system feature assessments for each condition (\sys vs. \baseline).
For TLX scores (see \cref{fig:nasa_tlx_combined_boxplot}), pairwise Wilcoxon signed-rank tests revealed no statistically significant differences on any of the measured dimensions: mental demand, temporal demand, performance, effort, and frustration, which suggests that subjective workload was comparable across the two conditions, despite the fact that,
as reported in \cref{sec:user_study_design}, most participants had limited prior experience with canvas-style tools, and \sys introduced a richer spatial interaction model than the sequential chat layout of \baseline.

\begin{figure*}[t]
    \centering
    \includegraphics[width=\linewidth]{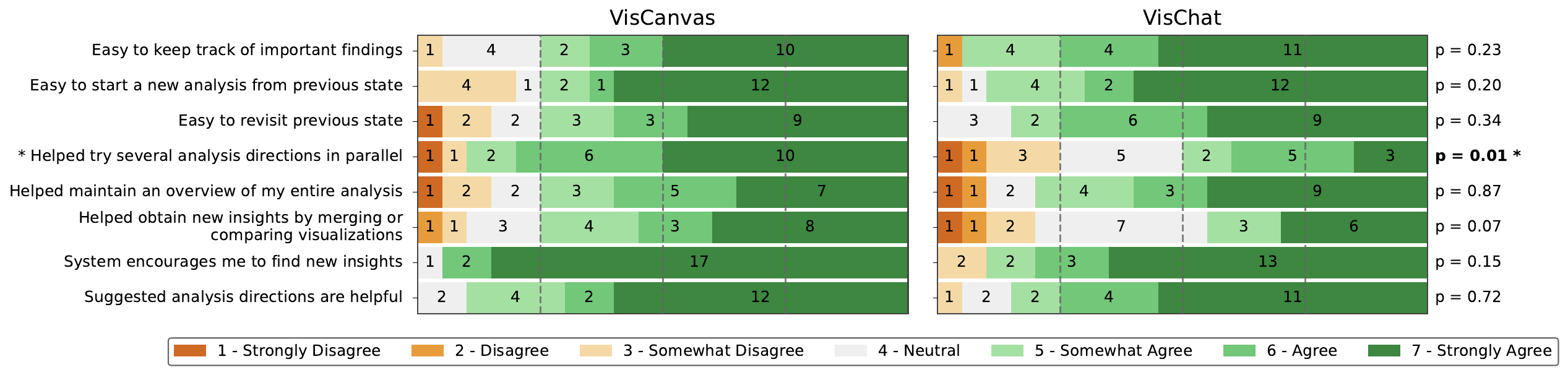}
    \caption{System Evaluation Results. Likert scale responses from the post-study survey. Wilcoxon signed-rank $p$-values are provided to the right.}
    \label{fig:system_evaluation}
\end{figure*}

For the system feature assessments (see \cref{fig:system_evaluation}), \sys received more favorable ratings than \baseline across nearly all assessed features. However, pairwise Wilcoxon signed-rank tests did not yield widespread statistical significance with Type I error controlled at the 0.05 level, with the notable exception that \sys received significantly higher ratings for supporting the exploration of multiple analysis directions in parallel. One possible explanation for the limited statistical differences is participants' familiarity with linear, chat-based systems: whereas \baseline followed a familiar sequential interaction model, \sys introduced a richer and less familiar spatial interaction model. Nevertheless, participants generally rated \sys's exploration-focused features favorably, without reporting significantly greater workload or frustration. These assessments are further contextualized by participants' freeform feedback in \cref{sec:user_preference_and_freeform_feedback}.

\subsubsection{User Preferences and Freeform Feedback (RQ2, RQ3)}
\label{sec:user_preference_and_freeform_feedback}
In the \textit{Post-Study Survey and Debrief} stage, participants were first asked about their preferences for \sys and \baseline for (i)~open-ended exploratory analysis tasks and for (ii)~closed, targeted analysis tasks (such as constructing a targeted chart to directly answer a hypothesis), using a 7-point Likert scale where 1 signifies a preference for \baseline and 7 indicates a preference for \sys.

For open-ended analysis, responses were highly skewed toward \sys ($\mu = 5.35, \sigma = 2.22$): 14/20 participants rated above the midpoint, and 11/20 participants selected the strongest preference for \sys (a score of 7).
In contrast, preferences for the closed and targeted analysis tasks were more mixed ($\mu= 3.8, \sigma = 2.16$):
8/20 participants preferred \sys, 10/20 preferred \baseline, and 2/20 selected the midpoint (indicating no clear overall preference). 

Freeform feedback shed further light on how participants preferred to use and benefit from the tools, such as the features and interactions considered beneficial or detrimental to success. To analyze this feedback, we conducted a thematic analysis, which is summarized into the following high-level findings.

\textbf{Finding 1 [C1]: Spatial visibility and semantic zoom supported navigation and thought organization.}
Several participants emphasized that the canvas-based interface helped them navigate their analysis and keep track of the overall analysis flow.
P3 noted, \textit{``VisCanvas is much more open than VisChat, and I can have a better view of where I have been and where I am going. In VisChat I had to scroll through the chat history to see what I had done.''} Similarly, P11 commented on \sys, \textit{``I can clearly see what I prompted and what the LLM generated.''}
In addition, P4 and P6 noted that the canvas-based interface helped them structure their thoughts in a ``\textit{tree format}'' while keeping past visualizations visible and organized for quick reference and easy access. P9 also highlighted semantic zoom as useful ``\textit{to focus on certain visualizations.}''
However, two participants suggested making the zoom level lockable and manually toggleable as a way for more fine-grained control.

\textbf{Finding 2 [C2, C3]: Canvas-based interface encouraged parallel exploration.}
\sys's canvas-based interface promoted tracking and pursuing multiple analytical directions in parallel.
P9 explained, \textit{``VisCanvas helped map out my ideas and go into several different directions at once,''} and P2 discussed how, \textit{``with VisCanvas, we can have all possible suggestions and their visualizations on a single screen.''}
This result is further supported by the post-task survey (see \cref{fig:system_evaluation}): participants indicated that \sys significantly enhanced their ability to pursue multiple analysis directions simultaneously compared to \baseline ($p = 0.01$).

\textbf{Finding 3 [C3]: Comparison and merging encouraged new insights.}
Several participants mentioned \sys made it easier to merge and compare visualizations, which in turn encouraged them to ideate new directions and ideas.
For example, P18 noted, \textit{``having well-defined graphs sit next to each other for different questions was very useful for further analysis.''}
Relatedly, P16 described \textit{Fill Blank} as useful because it ``\textit{filled [the] gap between two charts,}'' helping users understand their connectivity and view related charts together.
This trend is in part supported by the post-task survey (see \cref{fig:system_evaluation}), where participants indicated that \sys helped them somewhat more than \baseline to obtain new insights by merging and comparing visualizations, though this difference did not reach statistical significance ($p = 0.07$).

\textbf{Finding 4: Familiarity and lower interaction overhead made chat-based interfaces appealing.}
Some participants preferred \baseline even for open-ended exploratory tasks, citing its lower interaction overhead, more familiar interaction style, and lower learning curve.
In the context of open-ended analysis, P16 explained,\textit{``VisChat would allow me to do that easily because of the familiar chat interface,''} and P7 similarly mentioned, \textit{``I would prefer VisChat because it was easier to use.''}
While our \textit{Post-Study Survey} preference results were 14/20 in favor of \sys for open-ended tasks, these comments highlight that, for at least some participants, there was a tension between the more complex user experience and feature set of \sys and the sequential, chat-based familiarity of \baseline.

\textbf{Finding 5: Linear focus and straightforward interaction supported targeted analysis.}
When reflecting on its potential for targeted analysis, participants more often described \baseline as better suited for pursuing predefined questions or validating hypotheses.
Several participants mentioned that its straightforward and intuitive interaction style matched a focused analysis flow.
For instance, P9 noted, \textit{``VisChat's interface is set up better to explore a single chain of thought compared to VisCanvas,''} and P19 commented, \textit{``I would prefer VisChat for targeted analysis since we can focus on one single chart at a time and, with a predefined question, I can focus on that.''}

\textbf{Finding 6 [C1]: Visibility and control over intermediate results remained valuable in targeted analysis.}
At the same time, a subset of participants preferred \sys in targeted exploration when they valued visibility and control across intermediate results.
P17 explained, \textit{``I can see most of my plots on screen at the same time rather than scrolling up a chat,''} and P12 appreciated its incremental workflow, noting that the step-by-step creation of graphs could be a ``\textit{game changer}'' in more complex analytical scenarios.

\subsubsection{User Strategies for Exploratory Visual Authoring (RQ4)}
\label{sec:user_strategies}

\begin{figure*}[t!]
    \centering
    \includegraphics[width=\textwidth]{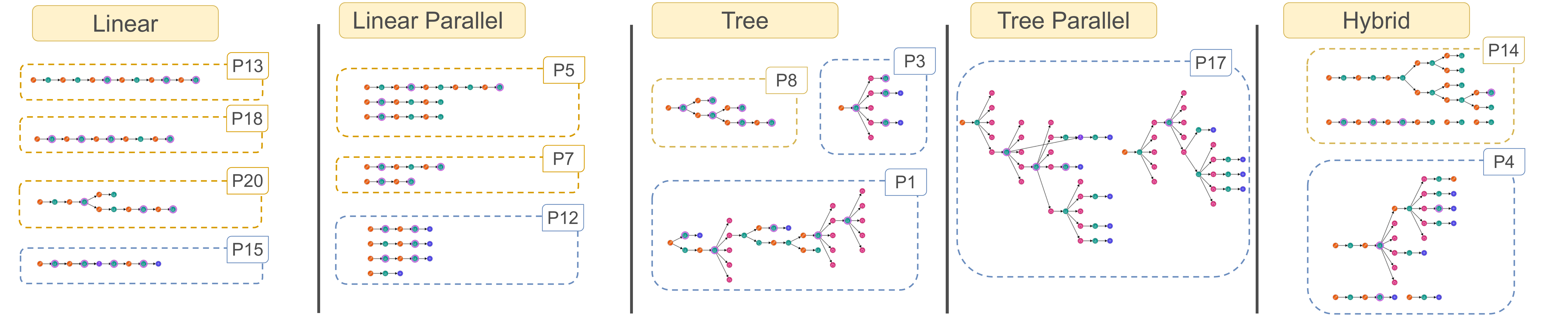}
    \caption{Representative examples of user exploration strategies, ordered by increasing structural complexity (left to right).
    Node colors correspond to their type: \iconcircle{text}{0.8em}{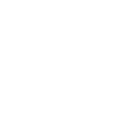} Text, \iconcircle{visualization}{0.8em}{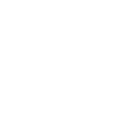} Visualization, \iconcircle{merge}{0.8em}{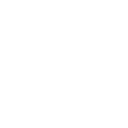} Merge, \iconcircle{suggestion}{0.8em}{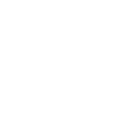} Suggest, and \iconcircle{menunode}{0.8em}{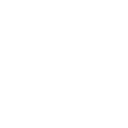} Menu nodes. \baseline (\colorcircle{VisChatColor} border) predominantly drove constrained, Linear explorations, whereas \sys ({\colorcircle{VisCanvasColor}} border) supported more divergent, complex behaviors, dominating the Hybrid category and generating the sole Tree Parallel structure.}
    \label{fig:glyph_overview}
\end{figure*}

System logs showed no significant between-interface differences in Visualization Nodes created during the \textit{Main Task} (\sys: $\mu = 8.95, \sigma = 3.56$; \baseline: $\mu = 7.95, \sigma = 4.26$) or unique variable sets explored (\sys: $\mu = 7.9$, $\sigma = 3.71$; \baseline: $\mu = 8.4$, $\sigma = 5.09$).
However, to better characterize the exploration and chart creation strategies of participants, we analyzed the topologies of their exploration structures (see \cref{fig:glyph_overview}). Based on a collaborative review by two researchers, we derived the following taxonomy of exploration patterns that were commonly performed by our participants:

\begin{itemize}[noitemsep]
    \item \textit{Linear:} A structure that is strictly or semantically linear.
    \item \textit{Linear Parallel:} Two or more concurrent linear explorations.
    \item \textit{Tree:} A singular hierarchical structure (i.e., one parent node with two or more child nodes).
    \item \textit{Tree Parallel:} Two or more concurrent tree structures.
    \item \textit{Hybrid:} Any combination of tree and linear structures.
\end{itemize}

To classify these structures, one author developed a codebook from the taxonomy, and three authors independently coded all participant topologies, achieving 82.5\% initial agreement; disagreements were resolved through discussion. The complete set of participant topologies is provided in Appx.~F
alongside visualizations of the distributions.

Data classification revealed distinct distributional trends when contrasting the two interfaces. \sys showed a clear preference in \textit{Hybrid} exploration (9 vs. 4 for \baseline) and \textit{Tree}-based exploration (5 vs. 2). Conversely, \textit{Linear} exploration was heavily favored in \baseline (8 vs. 1). These results align with our initial notions regarding the navigational affordances of each interface, and support our goal of exploratory analyses via \sys.

Only one interface, \sys, was used for the \textit{Tree Parallel} topology (only created once). For \textit{Linear Parallel} exploration, the counts were similar (4 for \sys vs. 6 for \baseline), which we consider comparable given the sample size. Finally, exploration approaches were consistent across datasets, with at most a one-participant difference per category.

A closer inspection of target node types at branching points further clarifies how \sys's features contributed to these exploration structures. In this analysis, we counted \textit{Suggest} nodes only when the suggested prompt was subsequently connected to a Visualization Node, thereby excluding suggestions that were generated but not pursued. Among branch targets that led to visualization results, Visualization Nodes accounted for the largest share (42/91, 46.2\%), followed closely by adopted \textit{Suggest} nodes (37/91, 40.7\%). \textit{Modify} nodes accounted for a smaller share (7/91, 7.7\%), while \textit{Fill Blank} (2/91, 2.2\%), \textit{Merge} (2/91, 2.2\%), and \textit{Duplicate} (1/91, 1.1\%) were comparatively rare as branch targets. These results suggest that \textit{Suggest} played a substantial role in supporting branching exploration, while branching also emerged through the \textit{Branch} operation in Edge Dropdown Menus and the \textit{Modify} operation in Menu Nodes.

\subsubsection{LLM Model Usage \& Performance (RQ5)}
\label{sec:performance_results}

Finally, to evaluate the system performance of \sys, we analyzed the response times of all visualization generation requests made during the study sessions.
Across 20 participants, \sys received 291 visualization generation requests, each of which triggered the visualization generation pipeline described in \cref{sec:backend}.
\texttt{gpt-4.1-nano} handled the majority of requests ($n = 261$, 89.7\%), while \texttt{gpt-5-nano} was used for the remaining requests ($n = 30$, 10.3\%), selected by 10/20 participants for a subset of their interactions.
The mean end-to-end response time was $6.70$s ($\sigma = 4.08$s) for \texttt{gpt-4.1-nano} and $32.78$s ($\sigma = 15.61$s) for \texttt{gpt-5-nano}.
The latency of \texttt{gpt-4.1-nano} was sufficiently low to maintain interactive responsiveness, while the longer response times of \texttt{gpt-5-nano} reflect the greater reasoning demands of complex specification generation tasks.
We define success rate as the proportion of generated Vega-Lite specifications that compiled without errors; \texttt{gpt-4.1-nano} achieved 96.2\% and \texttt{gpt-5-nano} achieved 96.7\%, indicating consistently reliable output across both models.

Participants' open-ended feedback indicated that waiting times were generally acceptable.
P7 noted, \textit{``I was primarily using the gpt-4.1-nano which generated the results fast.''}
Even P16, who used \texttt{gpt-5-nano} on several occasions, mentioned, \textit{``gpt-5-nano obviously took some more time, but overall the LLM timing was okay.''}

While waiting for LLM responses, two distinct behavioral patterns were observed: passively waiting while focusing on the anticipated output, or actively exploring the data and system features in parallel.
P10 noted, \textit{``I just thought about whether it would generate the correct visualization or not,''} and P11 similarly mentioned, \textit{``I was just wondering if the graph would look how I had pictured it.''}
In contrast, P6 mentioned, \textit{``I was thinking about which other charts could be generated while I was waiting for the LLM's response,''} and P18 noted, \textit{``While I was waiting, I simply did something else on the system since I was still exploring all the system's features.''}
These observations indicate that participants did not uniformly wait idly, but rather that some actively leveraged wait time to pursue parallel analyses or explore system functionality.



\section{Discussion}
\sys is an LLM-powered node-based exploratory visualization authoring tool that externalizes intermediate visualization states, branching paths, and relationships among analytical directions.
Our evaluation results suggest that the majority of participants preferred \sys over the baseline chat-based interface, particularly for open-ended analysis tasks, and also that
\sys supports the design considerations in \cref{sec:design_challenges_goals} (\textbf{C1}--\textbf{C3}) by making exploration structures visible, supporting branching and comparison, and helping users move between broad exploration and focused refinement.
More specifically, \textbf{C1} is supported by Finding~1 and the topology analysis: participants described the canvas and semantic zoom as useful for seeing where they had been and where they could go next, and \sys produced more Tree and Hybrid exploration structures than \baseline.
\textbf{C2} is supported by Finding~2 and the topology analysis: participants reported that \sys helped them view possible suggestions and pursue several directions at once, and adopted \textit{Suggest} prompts formed a substantial share of branch targets that led to visualizations. Because \baseline used the same suggestion backend, this difference cannot be attributed to higher suggestion quality in \sys. Instead, it suggests that \sys helped users preserve recommendations as persistent, spatially organized alternatives for parallel exploration.
\textbf{C3} is supported by Findings~2, 3 and the topology analysis, which together show that participants used \sys to pursue multiple directions in parallel and to compare or merge visualization states into new analysis directions.
Below, we reflect on several lessons learned from our work and discuss opportunities for future work in this area.

\textbf{Current System Limitations.}
The current system uses a data summary rather than the entire dataset to generate analysis goals and Vega-Lite specifications.
This approach was chosen to prioritize response time, though it may limit the accuracy of the generated visualization specifications or analysis goal suggestions, particularly for large or complex datasets.
Although no participant reached the model context-window limit or encountered context-length errors in our study, prompt size may increase for datasets with many columns, long root-to-node analysis paths used by the \textit{Suggest} operation, or complex Vega-Lite specifications passed as graph context. Future work could address this limitation through explicit token-budget monitoring, context pruning, or context compaction for large or complex analyses.
A related limitation is that \sys is not designed as a general-purpose multi-turn conversational agent. The current pipeline does not ask follow-up clarification questions before generating a visualization, nor does it support open-ended conversational exchanges such as discussing chart design alternatives without producing a Vega-Lite output. Future work could address this by introducing a new type of Text Input Node that supports multi-turn conversations with the agent.

\textbf{User Study Limitations.}
Our user study also has several limitations. Participants were all graduate computer science students, and many reported relatively high familiarity with data analysis and visualization tools, limiting the generalizability of our findings to novices, domain experts, and broader populations. Moreover, the 20-minute sessions and requirement to bookmark at least four insights may have constrained natural analysis strategies and limited our ability to observe longer-term behaviors such as revisiting branches or using provenance over extended sensemaking. The low NASA-TLX ratings further suggest that the tasks were manageable for these high-skill participants, potentially introducing ceiling/floor effects in task success, perceived difficulty, and workload; future evaluations should involve more diverse participants and longer, more challenging analysis tasks.

Beyond these limitations, we identified several promising directions for future work.

\textbf{Trade-offs of Node-based Exploration.}
Our results do not imply that node-based interfaces are universally preferable to linear chat workflows.
While \sys did not significantly increase NASA-TLX workload, several participants preferred \baseline for its familiarity, lower interaction overhead, and focused single-chain workflow, particularly for targeted analysis tasks.
These findings suggest that node-based provenance can introduce navigation and orientation costs as graphs grow, and future systems should consider a hybrid of both interfaces.

\textbf{Enriching the Design Space.}
Our current system focuses on Vega-Lite, as it efficiently supports visualization generation and modification.
However, this constrains the design space to chart types, data transformations, and interactions that fit within Vega-Lite's abstraction.
Future work could extend this approach to more expressive visualization grammars such as ECharts, Vega, or D3, which support richer custom interactions and more bespoke visual designs.

\textbf{Design-Oriented Visualization Authoring.}
Our work demonstrates that \sys effectively supports a branching and exploratory style of chart authoring.
In a downstream workflow, users could first use \sys to identify promising analytical branches and bookmark key insights, then reuse the generated charts and insights as starting points for further refinement in presentation-oriented design tools.
However, design-oriented visualization authoring also involves exploration: users may alternate between broad exploration and focused refinement~\cite{buxton_sketching_2007}, or design multiple alternatives in parallel~\cite{dow_parallel_2011, camburn_systematic_2015}.
Future systems could therefore build on our branching and comparison mechanisms to support such scenarios, including the construction and refinement of infographic designs, pictorial visualizations, or presentation-oriented visual storytelling, where the goal encompasses not only analytical correctness but also aesthetic quality and communicative effectiveness.

\textbf{Integrating Interaction-Augmented Instructions.}
Another promising direction is to enrich prompting with  interaction-augmented instructions~\cite{shen_prompting_2025}.
While our current system relies primarily on text input, multimodal inputs such as speech, sketching, or pointing gestures could help users express ambiguous intentions and high-level design goals more naturally.
Grounding such inputs in the provenance graph may improve the efficiency and interpretability of human-AI co-authoring.

\textbf{Collaborative Authoring.}
Finally, our provenance-based representation opens opportunities for collaborative visual authoring.
The graph can serve as a shared record of explored directions and analytical decisions, supporting asynchronous collaboration where collaborators resume work from meaningful intermediate states~\cite{park_storyfacets_2022}, as well as synchronous collaboration where multiple users can inspect, branch, and discuss visualization states in real time, drawing on interaction patterns from tools such as Miro or Figma.

\section{Conclusion}

In this work, we introduced \sys, a node-based interface for exploratory visualization authoring with LLMs.
Motivated by the mismatch between exploratory visual analysis workflows and prevailing linear chat interfaces, \sys integrates visualization authoring with exploration provenance on a non-linear canvas, where Vega-Lite visualization states can be created, revised, branched, compared, and merged.
\sys combines a graph-centric interface with an LLM pipeline for generating visualizations and context-aware suggestions, helping users maintain multiple analytical directions while moving between focused refinement and breadth-oriented exploration.
Our controlled study with 20 participants demonstrates that \sys promotes more diverse exploratory authoring behaviors and is preferred for open-ended analysis, while achieving levels of cognitive load and usability comparable with those of a chat-based baseline.
Collectively, our work offers an initial step toward LLM-assisted visualization authoring environments that support complex, non-linear exploratory analysis workflows.

\clearpage
\bibliographystyle{abbrv-doi-hyperref}

\bibliography{references, others}

\clearpage
\appendix
\crefalias{section}{appendix}
\crefalias{subsection}{appendix}
\section{Prompts for Visualization Generation}
In each prompt, certain variables are enclosed within curly brackets and colored blue for easy recognition.

\subsection{Prompt for Analysis Goal Generator}
\label{appendix:analysis_goal_generator_prompts}
The following variables are used in this prompt:
\begin{itemize}
    \item \texttt{state['data\_summary']}: Data summary generated by Data Summarizer.
    \item \texttt{state['vis\_instruction']}: Visualization instruction from the user.
    \item \texttt{state['input\_vega\_specs']}: Existing Vega-Lite specifications from upstream nodes (if present).
\end{itemize}

\begin{codeboxframe}
\begin{lstlisting}[style=codebox]
    You are an experienced data analyst skilled in generating analysis goals.

    # Task
    Generate an analysis goal given the Data Summary, Visualization Instruction, and the Existing Vega-Lite Specifications (if provided).

    # Input
    - Data Summary: {json.dumps(state['data_summary'], indent=2)}
    - Visualization Instruction: {state['vis_instruction']}
    {f"- Existing Vega-Lite Specifications: {json.dumps(state['input_vega_specs'], indent=2)}" if state.get('input_vega_specs') else ""}

    # Output
    Return ONLY a string that represents the analysis goal.
    Do not include any explanation, markdown formatting, or code blocks. Just return the raw string.
 \end{lstlisting}
\end{codeboxframe}

\subsection{Prompt for Vega-Lite Spec Generator}
\label{appendix:vega_spec_generator_prompts}
The following variables are used in this prompt:
\begin{itemize}
    \item \texttt{vega\_lite\_template}: A generic Vega-Lite template with \texttt{<stub>} fields to be filled by the LLM.
    \begin{codeboxframe}
        \begin{lstlisting}[style=jsoncode,gobble=12]
            {
              "$schema": "https://vega.github.io/schema/vega-lite/v6.json",
              "title": "<stub>",
              "description": "<stub>",
              "data": {
                "url": "$data_url"
              },
              "transform": "<stub>",
              "mark": "<stub>",
              "encoding": "<stub>"
            }
        \end{lstlisting}
    \end{codeboxframe}
    \item \texttt{vega\_lite\_examples}: Example Vega-Lite specifications for few-shot prompting.
    \item \texttt{state['data\_summary']}: Data summary generated by Data Summarizer.
    \item \texttt{state['analysis\_goal']}: Analysis goal generated by Analysis Goal Generator.
    \item \texttt{state['vis\_instruction']}: Visualization instruction from the user. For \textit{Merge} and \textit{Fill Blank}, the system uses the following predefined instructions.
        \begin{itemize}
            \item \textit{Merge}
                \begin{codeboxframe}
                \begin{lstlisting}[style=codebox]
    Integrate the two visualizations into a single comparative view. Identify shared attributes (e.g., common axis fields) and align them accordingly. If a shared axis exists, generate a layered or dual-axis chart to support direct comparison; if using a layered chart, limit it to at most two layers to keep the chart readable. Otherwise, construct a visualization that meaningfully relates the two sources (e.g., a repeated chart) to reveal potential relationships or contrasts.
                \end{lstlisting}
                \end{codeboxframe}
        \item \textit{Fill Blank}
                \begin{codeboxframe}
                \begin{lstlisting}[style=codebox]
    Generate an intermediate visualization that complements and connects the two given visualizations. Identify shared or related attributes, differences in analytical focus, and potential causal or correlational relationships. Create a visualization that fills the analytical gap, making the reasoning between the first and second visualization more explicit and interpretable.
                \end{lstlisting}
                \end{codeboxframe}
        \end{itemize}
    \item \texttt{state['input\_vega\_specs']}: Existing Vega-Lite specifications from upstream nodes (if present).
    \item \texttt{state['output\_vega\_spec']}: Generated Vega-Lite specification.
    \item \texttt{state['vega\_error\_message']}: Compile Error message.
\end{itemize}

\subsubsection{Vega-Lite Spec Generator}
\begin{codeboxframe}
\begin{lstlisting}[style=codebox]
    You are a helpful assistant highly skilled in writing PERFECT Vega-Lite specifications for visualizations. 

    # Task
    Fill in the `<stub>` sections of the Vega-Lite Template with the correct values to generate a visualization given the Data Summary, the Analysis Goal, Visualization Instruction (if provided), and Existing Vega-Lite Specifications (if provided).

    # Input
    - Vega-Lite Template: {vega_lite_template}
    - Data Summary: {json.dumps(state['data_summary'], indent=2)}
    - Analysis Goal: {state['analysis_goal']}
    {f"- Visualization Instruction: {state['vis_instruction']}" if state.get('vis_instruction') else ""}
    {f"- Existing Vega-Lite Specifications: {json.dumps(state['input_vega_specs'], indent=2)}" if state.get('input_vega_specs') else ""}
    - Vega-Lite Specification Examples: {json.dumps(vega_lite_examples, indent=2)}

    # Output
    Return ONLY a rationale and a valid JSON object that represents the complete Vega-Lite specification.
    Do not include any explanation, markdown formatting, or code blocks. Just return the raw JSON and a rationale string.
    The output should be in the following format:
    {{
        "rationale": "Brief explanation of design choices",
        "vega_spec": <complete vega-lite specification>
    }}

    ## Strict Rules
    - The specification you write MUST FOLLOW VISUALIZATION BEST PRACTICES i.e. meet the specified goal, apply the right transformation, use the right visualization type, use the right data encoding, and use the right aesthetics (e.g., ensure axis are legible). 
    - The transformations you apply MUST be correct and the fields you use MUST be correct. 
    - The visualization CODE MUST BE CORRECT and MUST NOT CONTAIN ANY SYNTAX OR LOGIC ERRORS (e.g., it must consider the field types and use them correctly). You can refer to the Vega-Lite Specification Examples to see how to write the specification.
    - You MUST first generate a brief plan for how you would solve the task e.g. what transformations you would apply e.g. if you need to construct a new column, what fields you would use, what visualization type you would use, what aesthetics you would use, etc. 
\end{lstlisting}
\end{codeboxframe}

\subsubsection{Reflection}
\begin{codeboxframe}
\begin{lstlisting}[style=codebox]
    You are an expert in fixing Vega-Lite specifications.

    # Task
    Fix the Vega-Lite specification to make it run successfully.

    # Input
    - Vega-Lite Specification: {state['output_vega_spec']}
    - Error Message: {state['vega_error_message']}

    # Output
    Return ONLY a valid JSON object that represents the fixed Vega-Lite specification.
    Do not include any explanation, markdown formatting, or code blocks. Just return the raw JSON.
    The output should be in the following format:
    {{
        "vega_spec": <fixed vega-lite specification>
    }}
\end{lstlisting}
\end{codeboxframe}

\section{Prompt for Analysis Goal Suggestion}
\label{appendix:analysis_goal_suggestion_prompts}
The following variables are used in this prompt:
\begin{itemize}
    \item \texttt{state['previous\_analysis\_goals']}: Previous analysis goals from upstream visualization nodes.
    \item \texttt{state['data\_summary']}: Data summary generated by Data Summarizer.
\end{itemize}

\begin{codeboxframe}
\begin{lstlisting}[style=codebox]
    You are an experienced data analyst skilled in generating analysis suggestions about data.
    
    # Task
    Generate 3-5 relevant next analysis steps as visualization instructions.
    Each suggestion should be a clear, actionable prompt for creating a visualization that:
      - Builds upon or complements the previous analysis goals
      - Explores new aspects not yet covered
      - Provides deeper insights or alternative perspectives

    # Input
    - Previous Analysis Goals: {json.dumps(state['previous_analysis_goals'], indent=2)}
    - Data Summary: {json.dumps(state['data_summary'], indent=2)}

    # Output
    Return ONLY a list of 3-5 suggestion strings.
    Do not include any explanation, markdown formatting, or code blocks. Just return the raw list.

    # Guidelines
    - DO NOT repeat analysis goals that have already been explored
    - Suggest analyses that naturally follow from the previous analysis goals
    - Consider both broader explorations and deeper dives
    - Mix different visualization types and analysis methods
\end{lstlisting}
\end{codeboxframe}

\section{JSON Schemas and Graph Propagation Logic}
\label{appendix:graph_json_schemas}

Each node and edge within the \sys graph is represented as a JSON object. These schemas define critical properties, including node positioning, the Vega-Lite specification, the LLM-generated rationale, the analysis goal, and topological connections.

The following is an example JSON object representing a node:
\begin{codeboxframe}
\begin{lstlisting}[style=jsoncode,gobble=4]
    {
        "id": "ns4oNmeXUob3sOoKTKLch",
        "data": {
        "title": "Visualization",
        "status": "success",
        "icon": "BarChart",
        "visSpec": {
          "$schema": "https://vega.github.io/schema/vega-lite/v6.json",
          "title": "Annual Distribution of Top Spotify Tracks",
          "description": "Bar chart showing the number of top Spotify tracks released each year from 1998 to 2020.",
          "data": {
            "url": "/data/songs_small.csv"
          },
          "transform": [
            {
              "calculate": "datum.year",
              "as": "release_year"
            },
            {
              "aggregate": [
                {
                  "op": "count",
                  "as": "song_count"
                }
              ],
              "groupby": ["release_year"]
            }
          ],
          "mark": {
            "type": "bar",
            "tooltip": true
          },
          "encoding": {
            "x": {
              "field": "release_year",
              "type": "ordinal",
              "title": "Year"
            },
            "y": {
              "field": "song_count",
              "type": "quantitative",
              "title": "Number of Songs"
            }
          }
        },
        "analysisGoal": "Determine the annual distribution of top Spotify tracks by counting the number of songs released each year from 1998 to 2020.",
        "rationale": "To visualize the distribution of top Spotify tracks by year, a bar chart is most appropriate as it clearly shows the count of songs released each year. The data needs to be aggregated by year, counting the number of songs per year. The x-axis will represent the year, and the y-axis will represent the count of songs. Since the dataset spans from 1998 to 2020, the x-axis should be treated as a temporal or ordinal scale to accurately reflect the timeline. The bar height will correspond to the number of songs released in each year, providing a clear view of trends over time."
        },
        "width": 549,
        "height": 457,
        "position": {
            "x": 470.5,
            "y": 40
        },
        "type": "generate-vis-node",
        "measured": {
            "width": 549,
            "height": 457
        },
        "selected": false
    }
\end{lstlisting}
\end{codeboxframe}

The following is an example JSON object representing an edge:
\begin{codeboxframe}
\begin{lstlisting}[style=jsoncode,gobble=4]
    {
        "id": "ns4oNmeXUob3sOoKTKLch-AgNJTbtq_O98DHm-tzbGj",
        "source": "ns4oNmeXUob3sOoKTKLch",
        "target": "AgNJTbtq_O98DHm-tzbGj",
        "sourceHandle": "vis-generate-output",
        "targetHandle": "vis-placeholder-input",
        "type": "placeholder-edge"
    }
\end{lstlisting}
\end{codeboxframe}

The system supports reactive updates: re-running a \textit{Text Input Node} automatically propagates new LLM-generated specifications to all connected \textit{Visualization Nodes}. This chain-like propagation allows for the simultaneous revision of downstream leaf nodes. In the event of node deletion, the graph maintains continuity by reconnecting descendants to the deleted node's parent. However, because \textit{Text Input Nodes} cannot be directly linked to one another, any orphaned \textit{Text Input Nodes} resulting from a deletion are automatically removed.

\section{Backend System Flow}

\begin{figure*}[h]
    \centering
        \includegraphics[width=\linewidth]{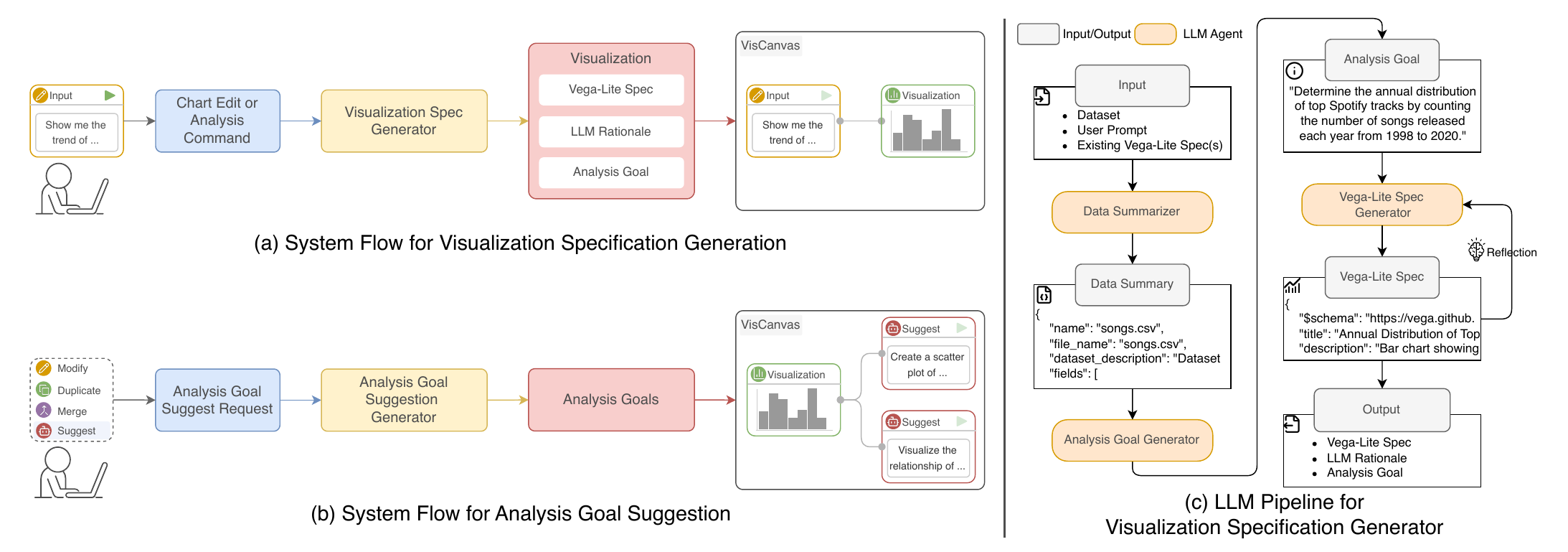}
    \caption{Backend system flow: (a) visualization specification generation; (b) analysis goal suggestion; (c) LLM pipeline for Visualization specification generator.}
    \label{fig:system_flow}
\end{figure*}

\Cref{fig:system_flow} provides an overview of the backend workflow, including visualization specification generation, analysis goal suggestion, and the LLM-based generation pipeline.

\section{Participant Demographics}
\label{appendix:participant_demographics}
\Cref{tab:participant_demographics} summarizes participants' demographic information and prior experience with LLMs, visualization, data analysis, online whiteboards, and visualization libraries.

\begin{table*}
\centering
\caption{Participant Demographics.}
\label{tab:participant_demographics}
\footnotesize
\begin{tabular}{llcccccccp{4.2cm}}
\toprule
& & & \multicolumn{5}{c}{Frequency} & \\
\cmidrule(lr){4-8}
ID & Age & Gender & LLM & LLM for Vis$^2$ & Vis & Data Analysis & Online Whiteboard tools & Vis Libraries \\
\midrule
P1  & 24 & Female & 5 & 3 & 4 & 4 & 2 & D3, ggplot2, Matplotlib \\
P2  & 25 & Male   & 5 & 2 & 2 & 2 & 2 & D3 \\
P3  & 25 & Male   & 5 & 4 & 4 & 4 & 3 & D3, Excel, Matplotlib, Tableau \\
P4  & 22 & Male   & 5 & 2 & 3 & 2 & -- & D3, Excel, Matplotlib \\
P5  & 21 & Male   & 5 & 2 & 5 & 4 & 2 & D3, Excel, Matplotlib, PowerBI, Seaborn \\
P6  & 22 & Female & 5 & 4 & 4 & 4 & 3 & D3, Excel, Matplotlib, Seaborn \\
P7  & 23 & Male   & 5 & 5 & 5 & 5 & 1 & D3, Excel, Matplotlib, PowerBI, Seaborn, Tableau \\
P8  & 22 & Male   & 5 & 2 & 4 & 2 & 2 & D3, Excel, Matplotlib \\
P9  & 25 & Female & 1 & 1 & 4 & 5 & 1 & D3, Matplotlib, Seaborn \\
P10 & 26 & Male   & 4 & 2 & 2 & 3 & 3 & D3, Excel, Matplotlib, PowerBI \\
P11 & 23 & Male   & 5 & 4 & 4 & 5 & 2 & D3, ggplot2, Matplotlib \\
P12 & 22 & Male   & 4 & 4 & 4 & 4 & 3 & D3, Matplotlib, WandB \\
P13 & 25 & Female & 4 & 2 & 4 & 4 & 4 & D3, Excel, Matplotlib, Seaborn \\
P14 & 23 & Male   & 4 & 3 & 5 & 4 & 3 & D3, Excel, Tableau \\
P15 & 23 & Female & 4 & 5 & 5 & 5 & 4 & D3, Matplotlib, PowerBI, Tableau \\
P16 & 22 & Female & 5 & 4 & 4 & 5 & 4 & D3, Matplotlib, PowerBI, Seaborn, Tableau \\
P17 & 25 & Male   & 4 & 2 & 5 & 5 & 3 & D3, Matplotlib, PowerBI, Seaborn \\
P18 & 22 & Female & 4 & 2 & 4 & 5 & 2 & D3, Excel, Matplotlib \\
P19 & 21 & Male   & 5 & 4 & 4 & 3 & 3 & D3, Excel, Matplotlib, PowerBI \\
P20 & 24 & Male   & 2 & 2 & 4 & 4 & 1 & D3 \\
\bottomrule
\end{tabular}
\vspace{2pt}
\begin{minipage}{\linewidth}
{\footnotesize $^1$Usage frequency scores: 5=At least once a day, 4=At least once a week, 3=At least once a month, 2=Less frequently but still occasionally, 1=Never. $^2$LLM usage for visualization authoring.}
\end{minipage}
\end{table*}

\clearpage
\section{Participants' Exploration Topologies}
\label{appendix:participant_topologies}
To code participants' exploration topologies, one author independently developed a codebook (reviewed and confirmed by the project team) using an open-coding method based on the taxonomy described in \cref{sec:user_strategies}. Three authors then coded the entire topology dataset, achieving an overall initial agreement of 82.5\%. The per-category agreements were: Linear (89\%), Linear Parallel (90\%), Tree (100\%), Tree Parallel (100\%), and Hybrid (62\%). The \textit{Hybrid} pattern was the most ambiguous class, with the lowest initial agreement, but all disagreements were resolved through discussion among the coders until consensus was reached across all topologies.

\begin{figure}[h]
    \centering
    \includegraphics[width=\columnwidth]{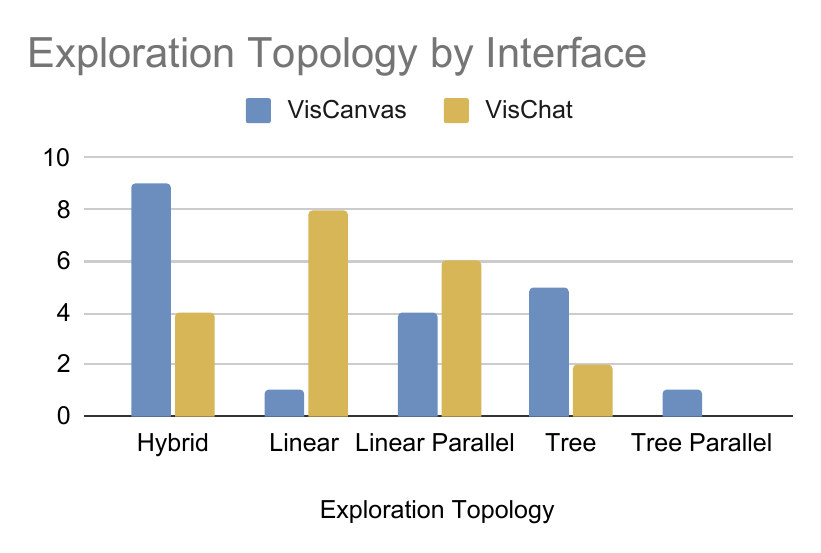}
    \caption{Distribution of participant topologies by interface. While \baseline performs comparably in linear exploration, \sys predominantly supports more divergent and complex behaviors, particularly in Hybrid and Tree Parallel structures.}
    \label{fig:exp_topologies_by_interface}
\end{figure}

The following figures illustrate individual participant topologies. Node colors and icons correspond to the node type:
\begin{description}
    \item[\iconcircle{text}{0.8em}{figs/icons/pen.png} Text (Modify)] Text Input Nodes, where users issue natural-language instructions to generate or revise visualizations.
    \item[\iconcircle{visualization}{0.8em}{figs/icons/chart.png} Visualization] Visualization Nodes that render Vega-Lite charts generated from user instructions.
    \item[\iconcircle{menunode}{0.8em}{figs/icons/plus.png} Menu] Menu Nodes that expose four operations: \textit{Modify}, \textit{Duplicate}, \textit{Merge}, and \textit{Suggest}.
    \item[\iconcircle{duplicate}{0.8em}{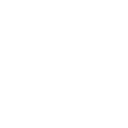} Duplicate] Duplicated Visualization Nodes that copy a parent visualization.
    \item[\iconcircle{merge}{0.8em}{figs/icons/merge.png} Merge] Merge Nodes that combine two Visualization Nodes into a single comparative visualization.
    \item[\iconcircle{suggestion}{0.8em}{figs/icons/robot.png} Suggest] Suggested Text Input Nodes (3--5 candidates) generated from accumulated analysis goals.
    \item[\iconcircle{fillblank}{0.8em}{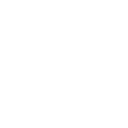} Fill Blank] Intermediate Visualization Nodes inserted between two connected Visualization Nodes.
\end{description}

\noindent\textit{Note: Visualization icons encircled by a \textbf{magenta circle} ({\colorcircle{magenta}}) indicate nodes that the participant bookmarked.}

\OverviewGraphic{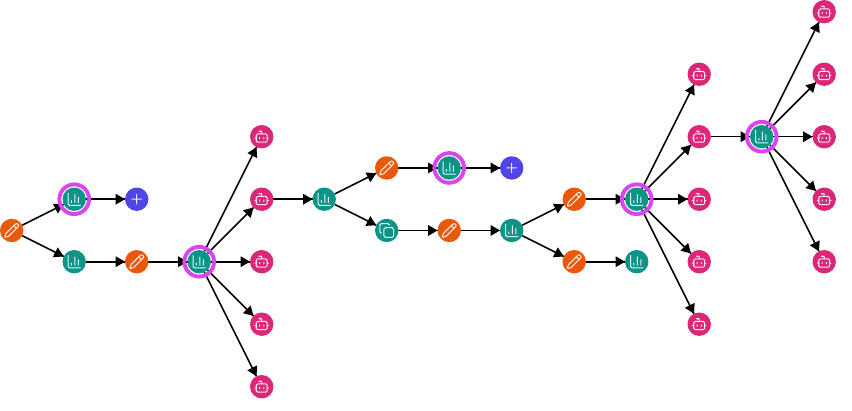}{1}{Social Media\&Mental Health}{VisCanvas}
\OverviewGraphic{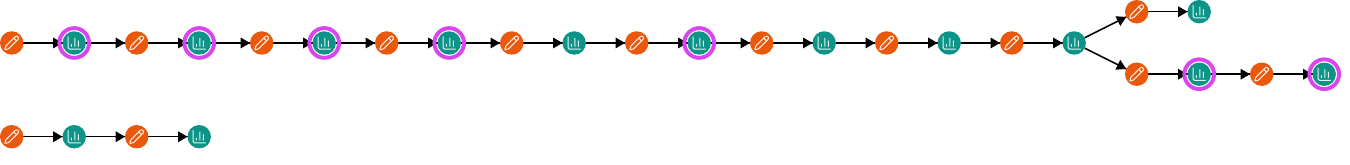}{1}{Spotify}{VisChat}
\OverviewGraphic{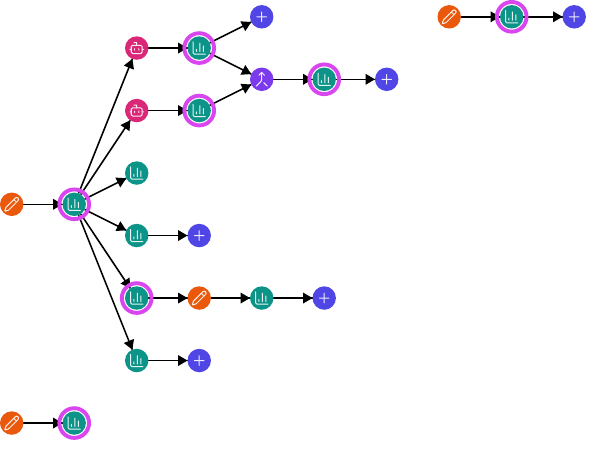}{2}{Spotify}{VisCanvas}
\OverviewGraphic{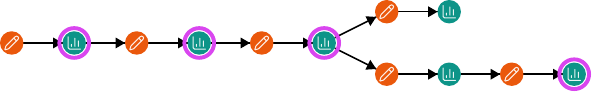}{2}{Social Media\&Mental Health}{VisChat}
\OverviewGraphic{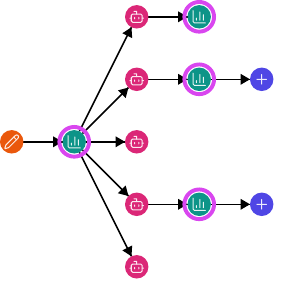}{3}{Spotify}{VisCanvas}
\OverviewGraphic{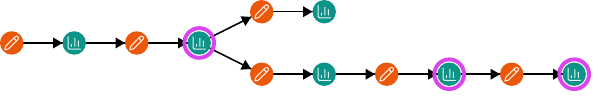}{3}{Social Media\&Mental Health}{VisChat}
\OverviewGraphic{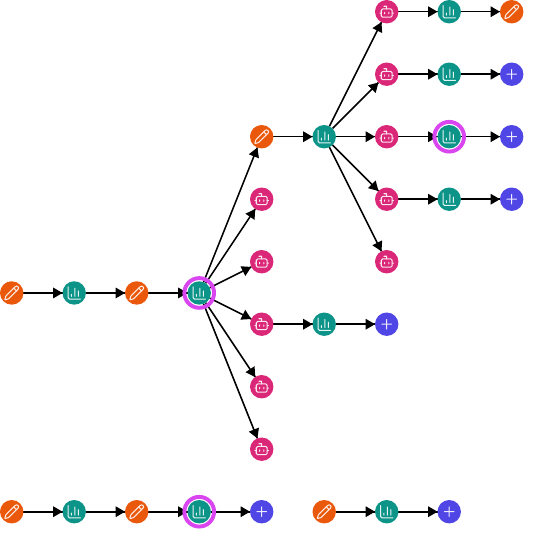}{4}{Social Media\&Mental Health}{VisCanvas}
\OverviewGraphic{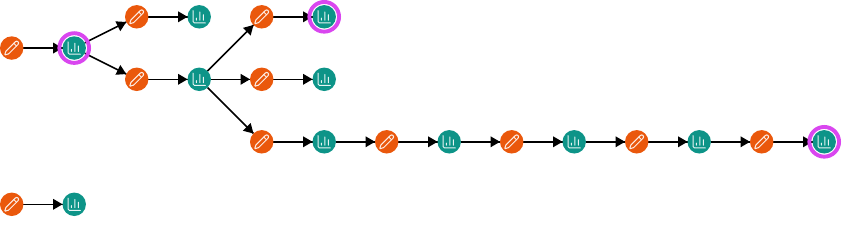}{4}{Spotify}{VisChat}
\OverviewGraphic{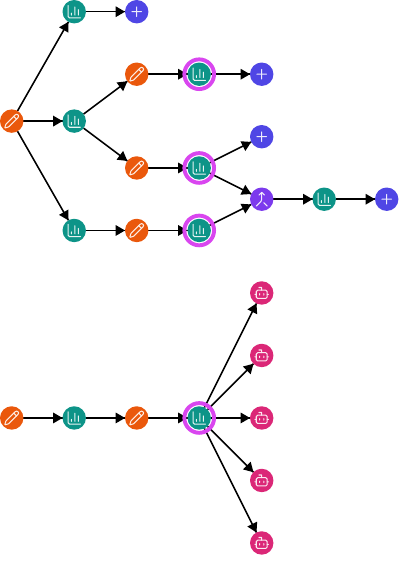}{5}{Social Media\&Mental Health}{VisCanvas}
\OverviewGraphic{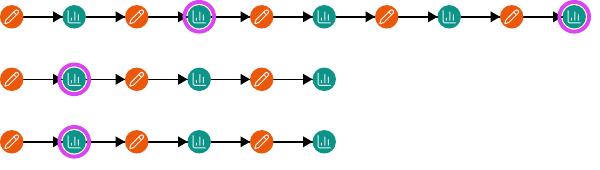}{5}{Spotify}{VisChat}
\OverviewGraphic{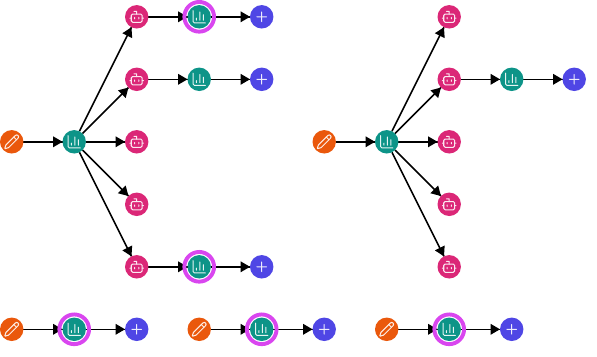}{6}{Spotify}{VisCanvas}
\OverviewGraphic{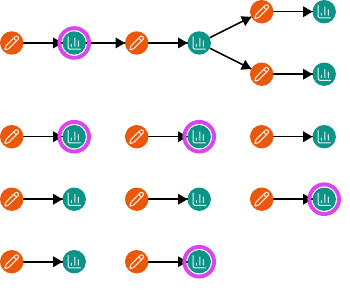}{6}{Social Media\&Mental Health}{VisChat}
\OverviewGraphic{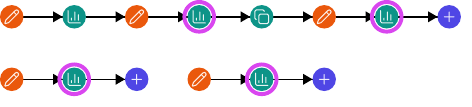}{7}{Spotify}{VisCanvas}
\OverviewGraphic{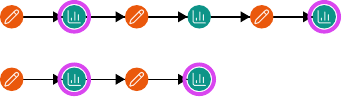}{7}{Social Media\&Mental Health}{VisChat}
\OverviewGraphic{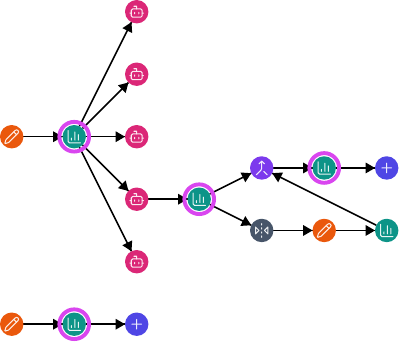}{8}{Social Media\&Mental Health}{VisCanvas}
\OverviewGraphic{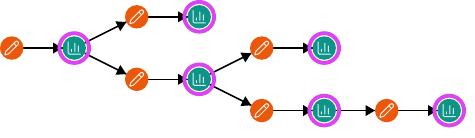}{8}{Spotify}{VisChat}
\OverviewGraphic{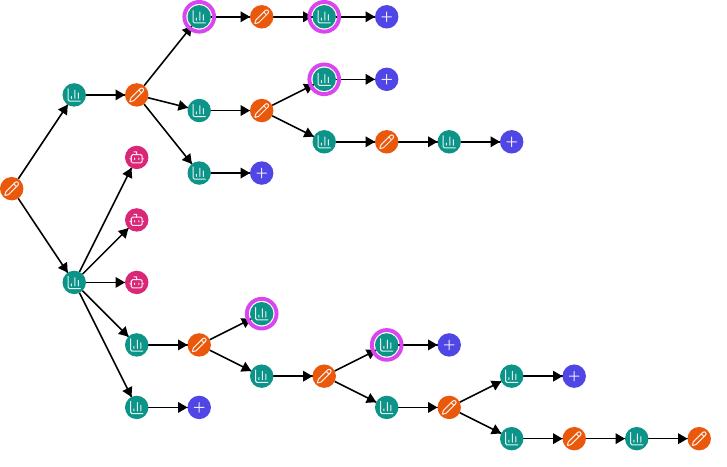}{9}{Social Media\&Mental Health}{VisCanvas}
\OverviewGraphic{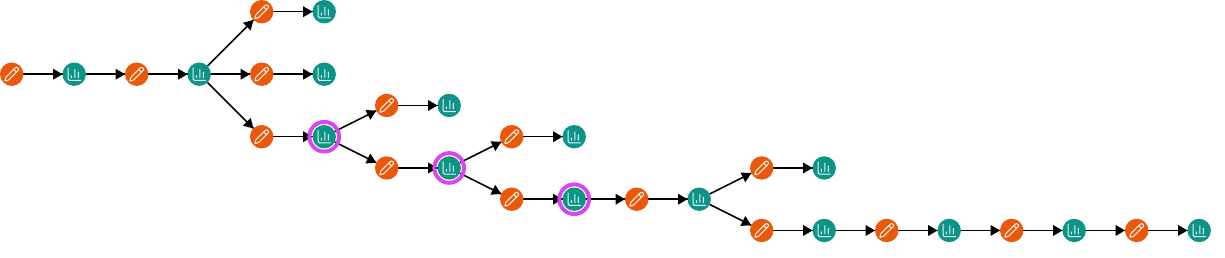}{9}{Spotify}{VisChat}
\OverviewGraphic{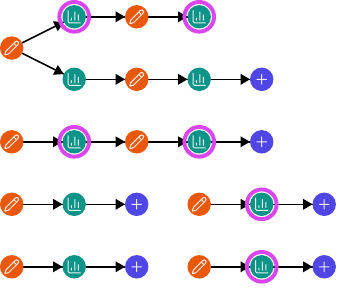}{10}{Spotify}{VisCanvas}
\OverviewGraphic{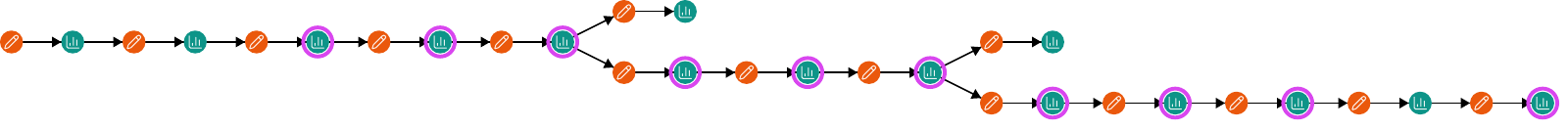}{10}{Social Media\&Mental Health}{VisChat}
\OverviewGraphic{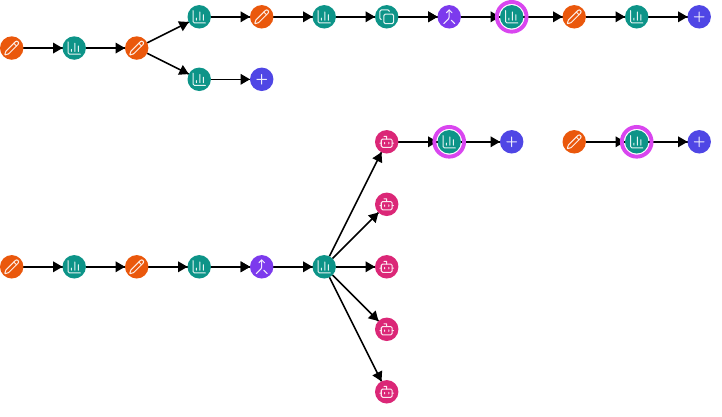}{11}{Spotify}{VisCanvas}
\OverviewGraphic{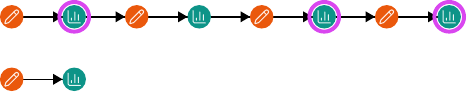}{11}{Social Media\&Mental Health}{VisChat}
\OverviewGraphic{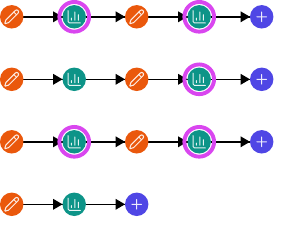}{12}{Social Media\&Mental Health}{VisCanvas}
\OverviewGraphic{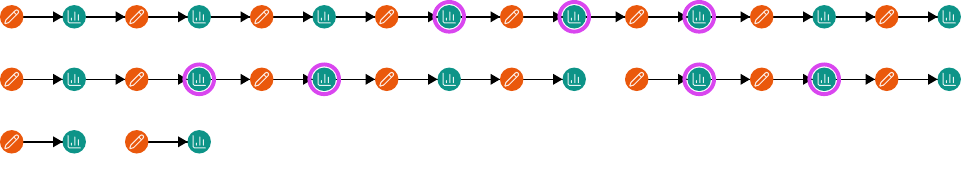}{12}{Spotify}{VisChat}
\OverviewGraphic{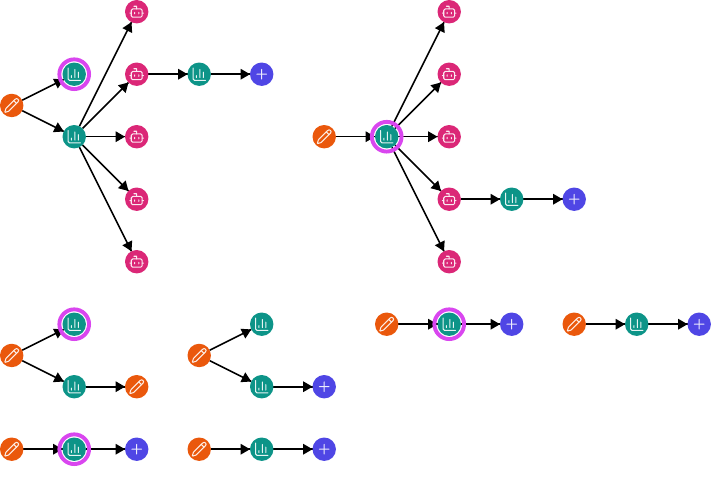}{13}{Social Media\&Mental Health}{VisCanvas}
\OverviewGraphic{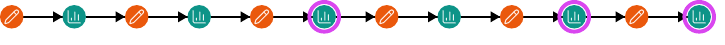}{13}{Spotify}{VisChat}
\OverviewGraphic{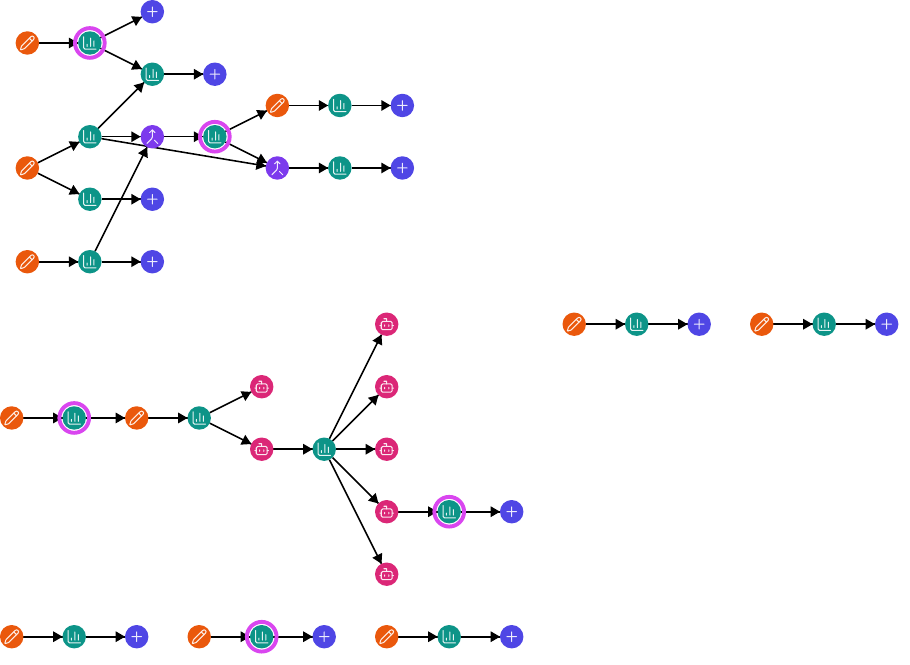}{14}{Spotify}{VisCanvas}
\OverviewGraphic{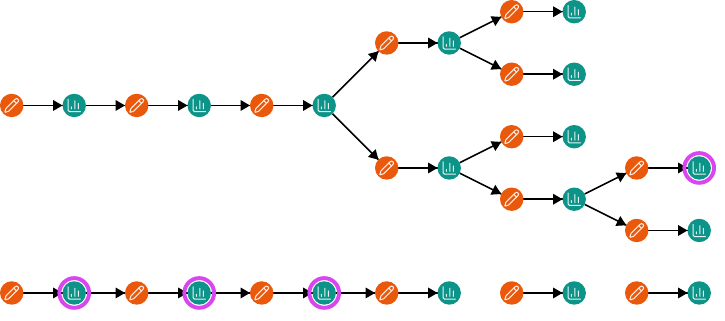}{14}{Social Media\&Mental Health}{VisChat}
\OverviewGraphic{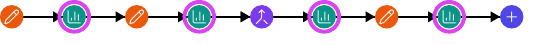}{15}{Spotify}{VisCanvas}
\OverviewGraphic{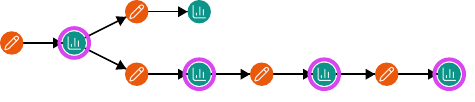}{15}{Social Media\&Mental Health}{VisChat}
\OverviewGraphic{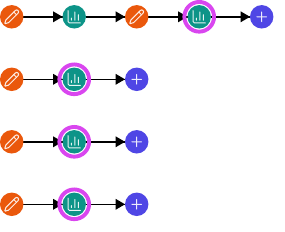}{16}{Social Media\&Mental Health}{VisCanvas}
\OverviewGraphic{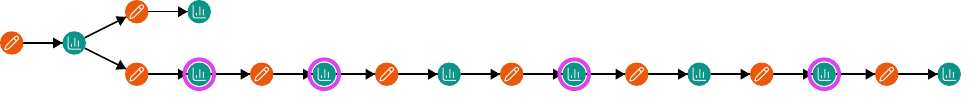}{16}{Spotify}{VisChat}
\OverviewGraphic{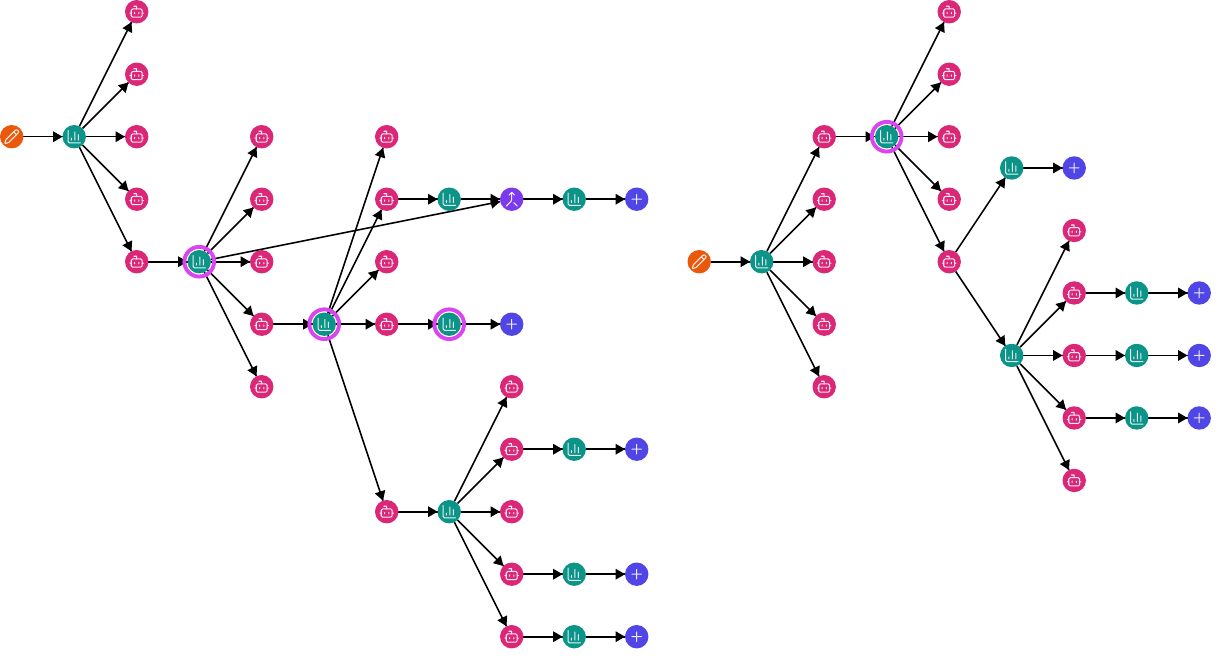}{17}{Social Media\&Mental Health}{VisCanvas}
\OverviewGraphic{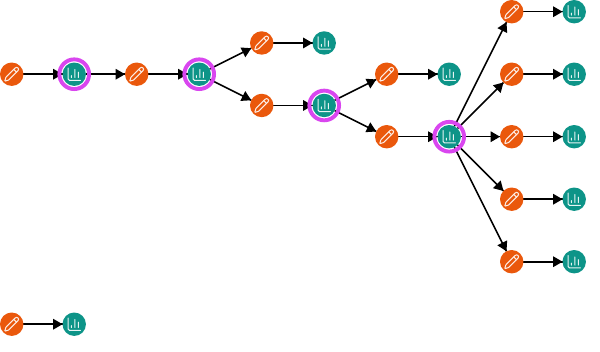}{17}{Spotify}{VisChat}
\OverviewGraphic{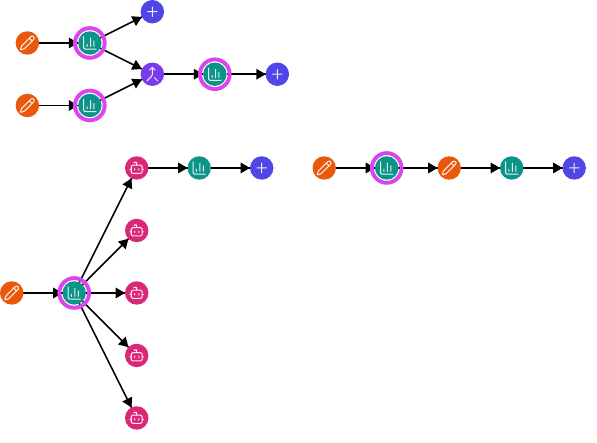}{18}{Spotify}{VisCanvas}
\OverviewGraphic{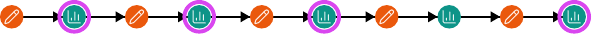}{18}{Social Media\&Mental Health}{VisChat}
\OverviewGraphic{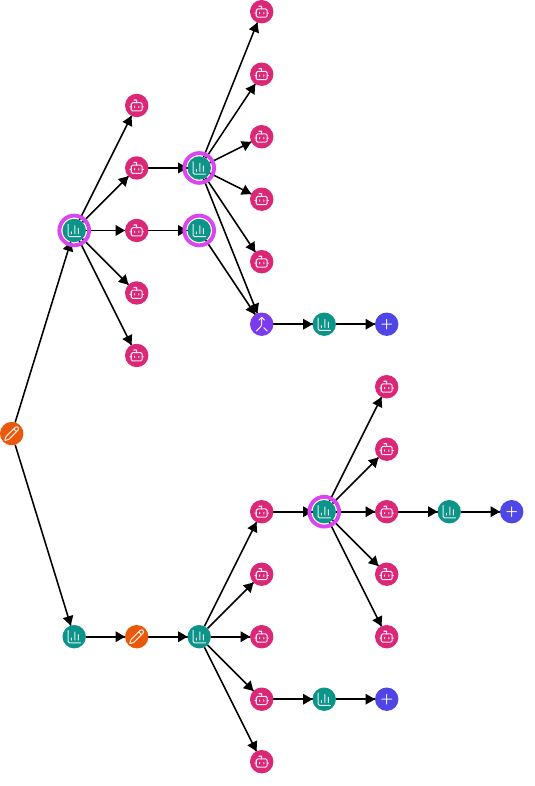}{19}{Spotify}{VisCanvas}
\OverviewGraphic{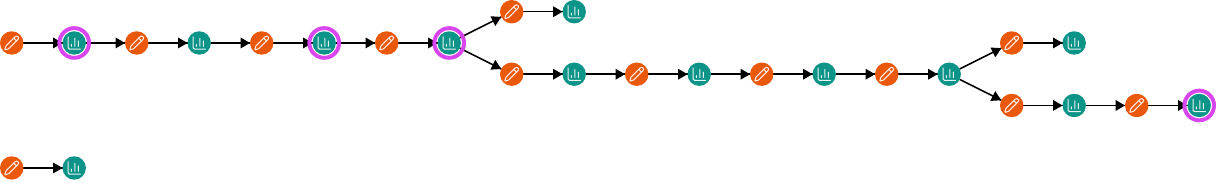}{19}{Social Media\&Mental Health}{VisChat}
\OverviewGraphic{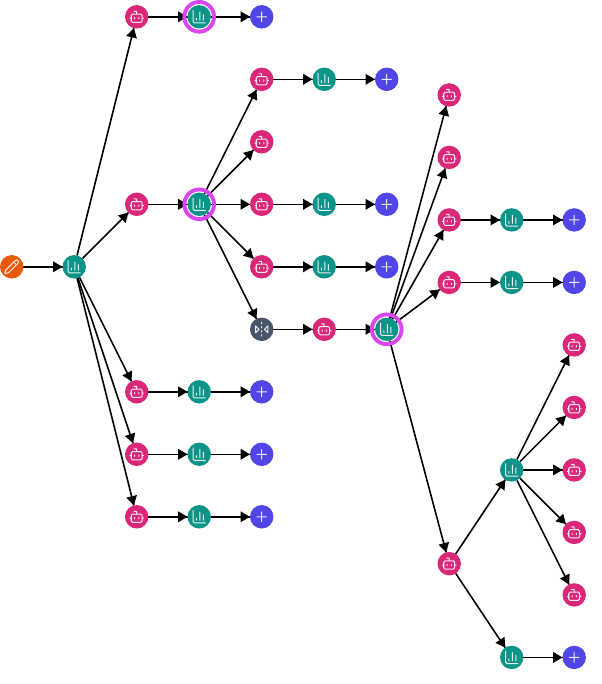}{20}{Social Media\&Mental Health}{VisCanvas}
\OverviewGraphic{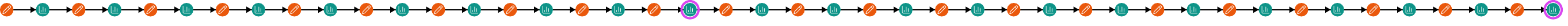}{20}{Spotify}{VisChat}





\end{document}